\documentclass[onefignum,onetabnum]{siamonline_171218/siamonline171218}
\usepackage{amsfonts,latexsym,verbatim,amscd,mathrsfs,color,array}
\usepackage{amsmath,amssymb,graphicx,color}
\usepackage{epstopdf}

\usepackage{subcaption}
\def\theequation{\thesection.\@arabic\c@equation}
\makeatother

\renewcommand{\theequation}{\thesection.\arabic{equation}}
\newtheorem{conjecture}{Conjecture}[section]
\newcommand{\R}{{\mathbb R}}

\numberwithin{equation}{section}
\title{An Asymptotic Analysis of Localized 3-D Spot Patterns for the
    Gierer-Meinhardt Model: Existence, Linear Stability and Slow Dynamics}

  \author{Daniel Gomez \thanks{Dept. of Mathematics, UBC, Vancouver,
      Canada. (corresponding author {\tt dagubc@math.ubc.ca})} \and
    Michael J. Ward \thanks{Dept. of Mathematics, UBC, Vancouver,
      Canada. {\tt ward@math.ubc.ca}} \and Juncheng Wei
    \thanks{Dept. of Mathematics, UBC, Vancouver, Canada. {\tt
        jcwei@math.ubc.ca}}}

\begin{document}

\maketitle

\begin{abstract}
  Localized spot patterns, where one or more solution components
  concentrates at certain points in the domain, are a common class of
  localized pattern for reaction-diffusion systems, and they arise in
  a wide range of modeling scenarios. Although there is a rather
  well-developed theoretical understanding for this class of localized
  pattern in one and two space dimensions, a theoretical study of such
  patterns in a 3-D setting is, largely, a new frontier.  In an
  arbitrary bounded 3-D domain, the existence, linear stability, and
  slow dynamics of localized multi-spot patterns is analyzed for the
  well-known singularly perturbed Gierer-Meinhardt (GM)
  activator-inhibitor system in the limit of a small activator
  diffusivity $\varepsilon^2\ll 1$.  Our main focus is to classify the different types of
  multi-spot patterns, and predict their linear stability properties,
  for different asymptotic ranges of the inhibitor diffusivity
  $D$. For the range $D={\mathcal O}(\varepsilon^{-1})\gg 1$, although
  both symmetric and asymmetric quasi-equilibrium spot patterns can be
  constructed, the asymmetric patterns are shown to be always
  unstable.  On this range of $D$, it is shown that symmetric spot
  patterns can undergo either competition instabilities or a Hopf
  bifurcation, leading to spot annihilation or temporal spot amplitude
  oscillations, respectively. For $D={\mathcal O}(1)$, only symmetric
  spot quasi-equilibria exist and they are linearly stable on
  ${\mathcal O}(1)$ time intervals. On this range, it is shown that
  the spot locations evolve slowly on an
  ${\mathcal O}(\varepsilon^{-3})$ time scale towards their
  equilibrium locations according to an ODE gradient flow, which is
  determined by a discrete energy involving the reduced-wave Green's
  function. The central role of the far-field behavior of a certain
  core problem, which characterizes the profile of a localized spot,
  for the construction of quasi-equilibria in the $D={\mathcal O}(1)$
  and $D={\mathcal O}(\varepsilon^{-1})$ regimes, and in establishing
  some of their linear stability properties, is emphasized. Finally,
  for the range $D={\mathcal O}(\varepsilon^{2})$, it is shown that
  spot quasi-equilibria can undergo a peanut-splitting instability,
  which leads to a cascade of spot self-replication
  events. Predictions of the linear stability theory are all
  illustrated with full PDE numerical simulations of the GM model.	
\end{abstract}

\section{Introduction}\label{sec:introuction}

We analyze the existence, linear stability, and slow dynamics of
localized $N$-spot patterns for the singularly perturbed
dimensionless Gierer-Meinhardt (GM) reaction-diffusion (RD) model
(cf.~\cite{gierer})
\begin{equation}\label{eq:pde_gm_3d}
   v_t = \varepsilon^2 \Delta v - v + \frac{v^2}{u}\,,\quad
   \tau u_t = D \Delta u - u + \varepsilon^{-2} v^2\,, \quad x\in\Omega\,;
   \qquad \partial_n v = \partial_n u = 0\,, \quad x\in\partial\Omega\,,
\end{equation}
where $\Omega\subset\mathbb{R}^3$ is a bounded domain,
$\varepsilon\ll 1$, and $v$ and $u$ denote the activator and inhibitor
fields, respectively. While the shadow limit in which
$D\rightarrow\infty$ has been extensively studied
(cf.~\cite{wei_surv}, \cite{wei_2014_book}, \cite{ward_2003_hopf}),
there have relatively few studies of localized RD patterns in 3-D with
a finite inhibitor diffusivity $D$ (see \cite{ren}, \cite{ei},
\cite{epstein}, \cite{tzou_2017_schnakenberg} and some references
therein). For 3-D spot patterns, the existence, stability, and
slow-dynamics of multi-spot quasi-equilibrium solutions for the
singularly perturbed Schnakenberg RD model was analyzed using
asymptotic methods in \cite{tzou_2017_schnakenberg}.  Although our
current study is heavily influenced by \cite{tzou_2017_schnakenberg},
our results for the GM model offer some new insights into the
structure of localized spot solutions for RD systems in
three-dimensions. In particular, one of our key findings is the
existence of two regimes, the $D={\mathcal O}(1)$ and
$D={\mathcal O}(\varepsilon^{-1})$ regimes, for which localized
patterns can be constructed in the GM-model, in contrast to the single
$D={\mathcal O}(\varepsilon^{-1})$ regime where such patterns occur
for the Schnakenberg model. Furthermore, our analysis traces this
distinction back to the specific far-field behaviour of the
appropriate core problem, characterizing the local behavior of a spot,
for the GM-model. By numerically solving the core problem, we
formulate a conjecture regarding the far-field limiting behavior of
the solution to the core problem. With the numerically established
properties of the core problem, strong localized perturbation theory
(cf.~\cite{ward_survey}) is used to construct $N$-spot
quasi-equilibrium solutions to \eqref{eq:pde_gm_3d}, to study their
linear stability, and to determine their slow-dynamics. We now give a
more detailed outline of this paper.

In the limit $\varepsilon\to 0$, in \S \ref{sec:quasi} we construct
$N$-spot quasi-equilibrium solutions to \eqref{eq:pde_gm_3d}. To do
so, we first formulate an appropriate core problem for a localized
spot, from which we numerically compute certain key properties of its
far field behavior. Using the method of matched asymptotic expansions,
we then establish two distinguished regimes for the inhibitor
diffusivity $D$, the $D={\mathcal O}(1)$ and
$D={\mathcal O}(\varepsilon^{-1})$ regimes, for which $N$-spot
quasi-equilibria exist. By formulating and analyzing a
nonlinear algebraic system, we then demonstrate that only symmetric
patterns can be constructed in the $D={\mathcal O}(1)$ regime, whereas
both symmetric and asymmetric patterns can be constructed in the
$D={\mathcal O}(\varepsilon^{-1})$ regime.

In \S\ref{sec:stability} we study the linear stability on an
${\mathcal O}(1)$ time scale of the $N$-spot quasi-equilibria
constructed in \S\ref{sec:quasi}. More specifically, we use the method
of matched asymptotic expansions to reduce a linearized eigenvalue
problem to a single globally coupled eigenvalue problem. We determine
that the symmetric quasi-equilibrium patterns analyzed in \S
\ref{sec:quasi} are always linearly stable in the $D={\mathcal O}(1)$
regime but that they may undergo both Hopf and competition
instabilities in the $D={\mathcal O}(\varepsilon^{-1})$
regime. Furthermore, we demonstrate that the asymmetric patterns
studied in \S \ref{sec:quasi} for the
$D={\mathcal O}(\varepsilon^{-1})$ regime are always unstable. Our
stability predictions are then illustrated in \S\ref{sec:examples}
where the finite element software FlexPDE6 \cite{flexpde} is used to
perform full numerical simulations of \eqref{eq:pde_gm_3d} for select
parameter values.

In \S\ref{sec:weak} we consider the weak interaction limit, defined by
$D={\mathcal O}(\varepsilon^2)$, where localized spots interact weakly
through exponentially small terms. In this regime,
\eqref{eq:pde_gm_3d} can be reduced to a modified core problem from
which we numerically calculate quasi-equilibria and determine their
linear stability properties. Unlike in the $D={\mathcal O}(1)$ and
$D={\mathcal O}(\varepsilon^{-1})$ regimes, we establish that spot
solutions in the $D={\mathcal O}(\varepsilon^2)$ regime can undergo
\textit{peanut-splitting} instabilities. By performing full numerical
simulations using FlexPDE6 \cite{flexpde}, we demonstrate that these
instabilities lead to a cascade of spot self-replication events in
3-D. Although spike self-replication for the 1-D GM model have been
studied previously in the weak interaction regime
$D={\mathcal O}(\varepsilon^2)$ (cf.~\cite{dp}, \cite{kww_split},
\cite{nishiura}), spot self-replication for the 3-D GM model has not
previously been reported.

In \S \ref{sec:gen_gm} we briefly consider the generalized GM system
characterized by different exponent sets for the nonlinear
kinetics. We numerically verify that the far-field behavior associated
with the new core problem for the generalized GM system has the same
qualitative properties as for the classical GM model
\eqref{eq:pde_gm_3d} This directly implies that many of the
qualitative results derived for \eqref{eq:pde_gm_3d} in \S
\ref{sec:quasi}--\ref{sec:slow_dynamics} still hold in this more
general setting.  Finally, in \S\ref{sec:discussion} we summarize our
findings and highlight some key open problems for
future research.

\section{Asymptotic Construction of an $N$-Spot Quasi-Equilibrium Solution}\label{sec:quasi}

In this section we asymptotically construct an $N$-spot
quasi-equilibrium solution where the activator is concentrated at $N$
specified points that are well-separated in the sense that
$x_1,\ldots,x_N\in\Omega$, $|x_i-x_j|={\mathcal O}(1)$ for $i\neq j$,
and $\text{dist}(x_i,\partial\Omega)={\mathcal O}(1)$ for
$i=1,\ldots,N$. In particular, we first outline the relevant core
problem and describe some of its properties using asymptotic and
numerical calculations. Then, the method of matched asymptotic
expansions is used to derive a nonlinear algebraic system whose
solution determines the quasi-equilibrium pattern. A key feature of
this nonlinear system, in contrast to that derived in
\cite{tzou_2017_schnakenberg} for the 3-D Schnakenberg model, is is
that it supports different solutions depending on whether
$D={\mathcal O}(1)$ or $D={\mathcal O}(\varepsilon^{-1})$.  More
specifically, we will show that the $D={\mathcal O}(1)$ regime admits
only spot quasi-equilibria that are symmetric to leading order,
whereas the $D={\mathcal O}(\varepsilon^{-1})$ regime admits both
symmetric and asymmetric $N$-spot quasi-equilibria.

\subsection{The Core Problem}\label{subsec:core_problem}

A key step in the application of the method of matched asymptotic
expansions to construct localized spot patterns is the study of the
core problem
\begin{subequations}\label{eq:core}
\begin{align}
  & \Delta_\rho V - V + U^{-1} V^2 = 0\,,\qquad \Delta_\rho U = -V^2\,,
    \qquad \rho>0\,, \label{eq:core_pde}\\
  \partial_\rho V(0) =  \partial_\rho & U(0)  = 0\,; \qquad
 V\longrightarrow 0\quad\text{and} \quad U \sim \mu(S) + {S/\rho}\,,
  \quad \rho\rightarrow\infty\,, \label{eq:core_bc}
\end{align}
\end{subequations}
where
$\Delta_\rho\equiv \rho^{-2}\partial_{\rho}\left[\rho^2\partial_\rho
\right]$.  For a given value of $S>0$, \eqref{eq:core} is to be solved
for $V=V(\rho;S)$, $U=U(\rho;S)$, and $\mu = \mu(S)$.  Specifying the
value of $S>0$ is equivalent to specifying the $L^2(\mathbb{R}^3)$
norm of $V$, as can be verified by applying the divergence theorem to
the second equation in \eqref{eq:core_pde} over an infinitely large
ball, which yields the identity
$S = \int_0^\infty \rho^2 \left[V(\rho)\right]^2\, d\rho$.

\begin{figure}[t!]
	\centering
	\begin{subfigure}[b]{0.33\textwidth}
		\centering
      		\includegraphics[width=\textwidth,height=4.2cm]{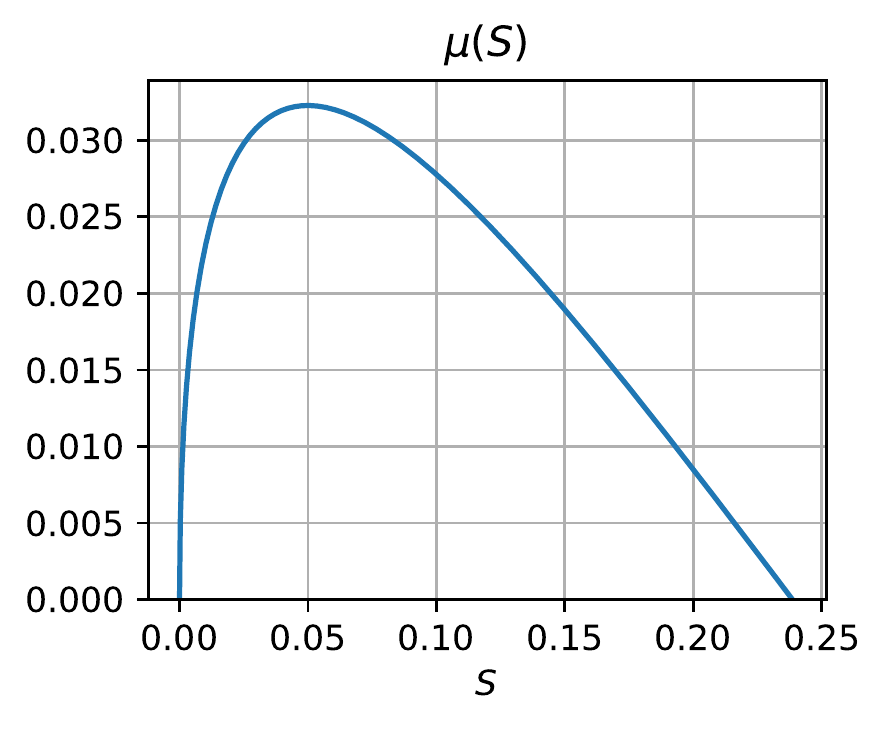}
		 \caption{}
                \label{fig:core_sol_mu}
	\end{subfigure}%
	\begin{subfigure}[b]{0.33\textwidth}
		\centering
		\includegraphics[width=\textwidth,height=4.2cm]{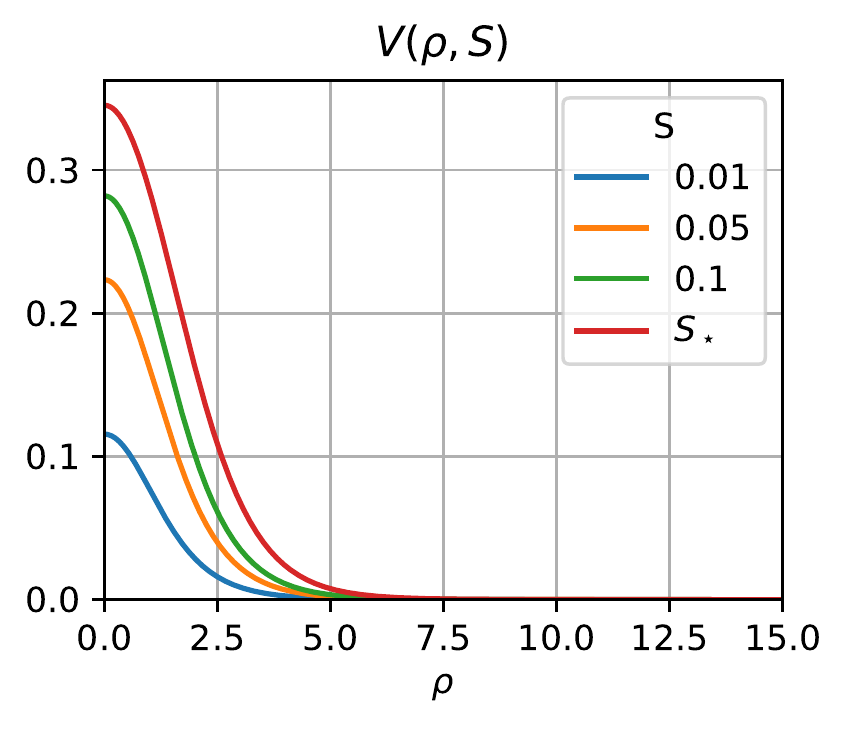}
		 \caption{}
                \label{fig:core_sol_activator}
	\end{subfigure}%
	\begin{subfigure}[b]{0.33\textwidth}
		\centering
		\includegraphics[width=\textwidth,height=4.2cm]{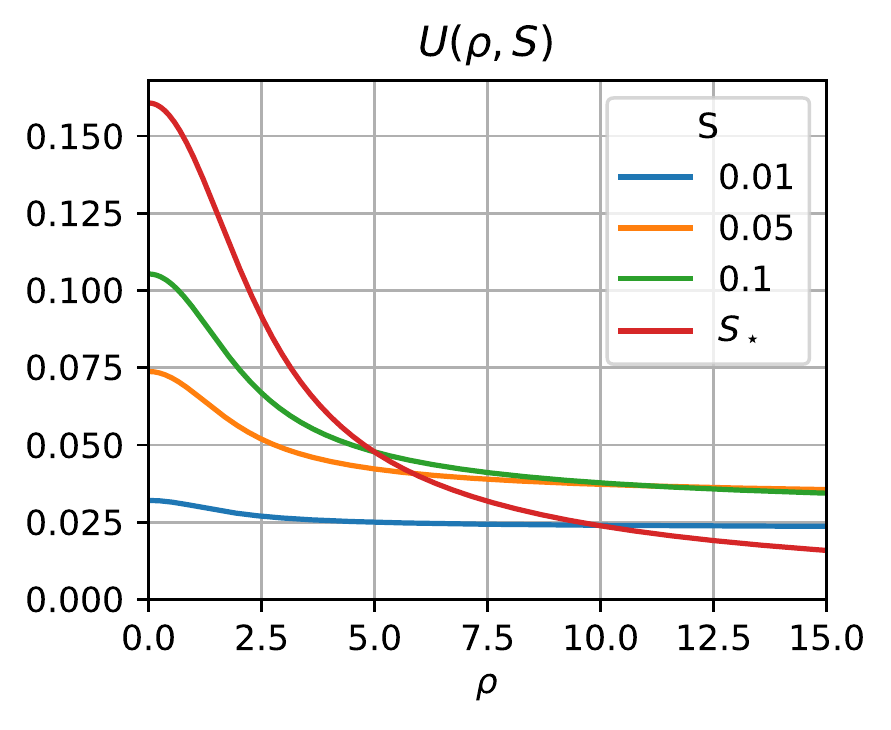}
		 \caption{}
                \label{fig:core_sol_inhibitor}
	\end{subfigure}%
	\caption{Plots of numerical solutions of the core problem
          \eqref{eq:core}: (a) $\mu(S)$ versus $S$, as well as the (b)
          activator $V$ and (c) inhibitor $U$, at a few select values of
          $S$. The value $S=S_\star \approx 0.23865$ corresponds to
          the root of $\mu(S)=0$.}
        \label{fig:core_sol}
\end{figure}

When $S\ll 1$ we deduce from this identity that
$V = {\mathcal O}(\sqrt{S})$. By applying the divergence theorem to
the first equation in \eqref{eq:core_pde} we get
$U={\mathcal O}(\sqrt{S})$, while from \eqref{eq:core_bc} we conclude
that $\mu={\mathcal O}(\sqrt{S})$. It is then straightforward to
compute the leading order asymptotics
\begin{equation}\label{eq:core_sol_small_S}
  V(\rho;S)\sim \sqrt{\frac{S}{b}}w_c(\rho)\,,\qquad
  U(\rho;S)\sim \sqrt{\frac{S}{b}}\,,\qquad \mu(S)\sim \sqrt{\frac{S}{b}}\,,
  \quad \mbox{for} \quad S\ll 1\,,
\end{equation}
where
$b \equiv \int_0^\infty \rho^2 \left[w_c(\rho)\right]^2\, d\rho \approx
10.423$ and $w_c>0$ is the unique nontrivial solution to
\begin{equation}\label{eq:homoclinic_equation}
  \Delta_\rho w_c - w_c + w_c^2 = 0\,,\quad \rho>0\,; \qquad
  \partial_\rho w_c(0) = 0\,,\qquad w_c\rightarrow 0\quad\text{as} \,\, 
  \rho\rightarrow\infty\,.
\end{equation}
We remark that \eqref{eq:homoclinic_equation} has been well studied,
with existence being proved using a constrained variational method,
while its symmetry and decay properties are established by a maximum
principle (see for example Appendix 13.2 of \cite{wei_2014_book}).
The limit case $S\ll 1$ is related to the \textit{shadow limit}
obtained by taking $D\rightarrow\infty$, for which numerous rigorous
and asymptotic results have previously been obtained
(cf.~\cite{wei_surv}, \cite{wei_2014_book}, \cite{ward_2003_hopf}).

Although the existence of solutions to \eqref{eq:core} have not been
rigorously established, we can use the small $S$ asymptotics given in
\eqref{eq:core_sol_small_S} as an initial guess to numerically
path-follow solutions to \eqref{eq:core} as $S$ is increased. The
results of our numerical computations are shown in Figure
\ref{fig:core_sol} where we have plotted $\mu(S)$, $V(\rho;S)$, and
$U(\rho;S)$ for select values of $S>0$. A key feature of the plot of
$\mu(S)$ is that it has a zero crossing at $S=0$ and
$S=S_\star \approx 0.23865$, while it attains a unique maximum on the
interval $0\leq S\leq S_\star$ at $S=S_\text{crit} \approx
0.04993$. Moreover, our numerical calculations indicate that
$\mu^{\prime\prime}(S)<0$ on $0<S\leq S_\star$. The majority of our subsequent
analysis hinges on these numerically determined properties of
$\mu(S)$. We leave the task of rigorously proving the existence of
solutions to \eqref{eq:core} and establishing the numerically verified
properties of $\mu(S)$ as an open problem, which we summarize in the
following conjecture:

\begin{conjecture}\label{conjecture}
  There exists a unique value of $S_\star>0$ such that \eqref{eq:core}
  admits a ground state solution with the properties that $V,U>0$ in
  $\rho>0$ and for which $\mu(S_\star) = 0$. Moreover, $\mu(S)$
  satisfies $\mu(S)>0$ and $\mu^{\prime\prime}(S)<0$ for all
  $0<S<S_\star$.
\end{conjecture}

\subsection{Derivation of the Nonlinear Algebraic System (NAS)}\label{quasi:match}

We now proceed with the method of matched asymptotic expansions to
construct quasi-equilibria for \eqref{eq:pde_gm_3d}.  First we seek an
inner solution by introducing local coordinates
$y = \varepsilon^{-1}(x-x_i)$ near the $i^\text{th}$ spot and letting
$v\sim D V_i(y)$ and $u\sim D U_i(y)$ so that the local steady-state
problem for \eqref{eq:pde_gm_3d} becomes
\begin{equation}
  \Delta_y V_i - V_i + U_i^{-1} V_i^2 = 0\,,\qquad \Delta_y U_i -
  \varepsilon^2 D^{-1} U_i + V_i^2 = 0\,,\qquad y\in\mathbb{R}^3\,.
\end{equation}
In terms of the solution to the core problem \eqref{eq:core} we
determine that
\begin{equation}\label{eq:sol_inner}
  V_i \sim V(\rho,S_{i\varepsilon}) + {\mathcal O}(D^{-1}\varepsilon^2)\,,\qquad
  U_i \sim U(\rho,S_{i\varepsilon}) + {\mathcal O}(D^{-1}\varepsilon^2)\,,
  \qquad \rho \equiv |y| = \varepsilon^{-1}|x-x_i|\,,
\end{equation}
where $S_{i\varepsilon}$ is an unknown constant that depends weakly on
$\varepsilon$. We remark that the derivation of the next order term
requires that $x_1,\ldots,x_N$ be allowed to vary on a slow time
scale.  This higher order analysis is done in \S
\ref{sec:slow_dynamics} where we derive a system of ODE's for the spot
locations.

To determine $S_{1\varepsilon},\ldots,S_{N\varepsilon}$ we now derive
a nonlinear algebraic system (NAS) by matching inner and outer
solutions for the inhibitor field. As a first step, we calculate in the
sense of distributions that
$\varepsilon^{-3} v^2 \longrightarrow 4\pi D^2 \sum_{j=1}^N
S_{j\varepsilon} \, \delta(x-x_j) + {\mathcal O}(\varepsilon^2)$ as
$\varepsilon\rightarrow 0^+$.  Therefore, in the outer region the
inhibitor satisfies 
\begin{equation}\label{eq:u_out}
  \Delta u - D^{-1} u = -4\pi\varepsilon D \sum_{j=1}^N S_{j\varepsilon}
  \delta(x-x_j) + {\mathcal O}(\varepsilon^3)\,,\quad x\in\Omega\,;
  \qquad \partial_nu=0\,,\quad x\in\partial\Omega\,.
\end{equation}
To solve \eqref{eq:u_out}, we let $G(x;\xi)$ denote the reduced-wave
Green's function satisfying
\begin{equation}\label{eq:greens_function_pde}
\begin{split}
  \Delta G - D^{-1} G &= -\delta(x-\xi)\,,\quad x\in\Omega\,;\qquad
  \partial_n G = 0\,,\quad x\in\partial\Omega\,, \\
  G(x;\xi) &\sim \frac{1}{4\pi|x-\xi|} + R(\xi) + \nabla_{x} R(x;\xi)
  \cdot (x-\xi) \,, \qquad \mbox{as} \quad x\to \xi \,,
\end{split}
\end{equation}
where $R(\xi)$ is the regular part of $G$. The solution to
\eqref{eq:u_out} can be written as
\begin{equation}\label{eq:u_outer}
  u \sim 4\pi \varepsilon D \sum_{j=1}^{N} S_{j\varepsilon} G(x;x_j) +
  {\mathcal O}(\varepsilon^3)\,.
\end{equation}

Before we begin matching inner and outer expansions to determine
$S_{1\varepsilon},\ldots,S_{N\varepsilon}$ we first motivate two
distinguished limits for the relative size of $D$ with respect to
$\varepsilon$. To do so, we note that when $D\gg 1$ the Green's
function satisfying \eqref{eq:greens_function_pde} has the regular
asymptotic expansion
\begin{equation}\label{eq:greens_function_large_D}
G(x,\xi) \sim D |\Omega|^{-1} + G_0(x,\xi) + {\mathcal O}(D^{-1})\,,
\end{equation}
where $G_0(x,\xi)$ is the Neumann Green's function satisfying
\begin{subequations}\label{eq:G0_equation}
\begin{align}
  \Delta G_0 &= \frac{1}{|\Omega|} -  \delta(x-\xi)\,, \quad x\in\Omega\,;
  \qquad \partial_n G_0 = 0\,,\quad x\in\partial\Omega\,;\qquad
          \int_{\Omega}G_0 \, dx = 0\,, \\
  G_0(x,\xi) &\sim \frac{1}{4\pi|x-\xi|} + R_0(\xi) + \nabla_{x}
    R_0(x;\xi) \cdot (x-\xi) \,, \qquad  \mbox{as}\quad x \to \xi \,,
\end{align}
\end{subequations}
and $R_0(\xi)$ is the regular part of $G_0$. In summary, for the two ranges
of $D$ we have
\begin{equation}\label{eq:greens_function_singular_behaviour}
  G(x,\xi) \sim \frac{1}{4\pi|x-\xi|} +
  \begin{cases} R(\xi) + o(1) \,, & D={\mathcal O}(1)\,, \\
    D|\Omega|^{-1} + R_0(\xi) + o(1) \,, & D\gg 1\,,\end{cases}\qquad \text{as}
  \quad |x-\xi|\rightarrow 0\,,
\end{equation}
where $R(\xi)$ is the regular part of $G(x,\xi)$. By matching the
$\rho\rightarrow\infty$ behaviour of $U_i(\rho)$ given by
\eqref{eq:sol_inner} with the behaviour of $u$ given by
\eqref{eq:u_outer} as $|x-x_i|\rightarrow 0$, we obtain in the
two regimes of $D$ that
\begin{equation}\label{eq:mu_all}
  \mu(S_{i\varepsilon}) = 4\pi\varepsilon \begin{cases} S_{i\varepsilon} R(x_i)
    + \sum_{j\neq i}S_{j\varepsilon} G(x_i,x_j)\,, & D={\mathcal O}(1)\,, \\
    S_{i\varepsilon} R_0(x_i) + \sum_{j\neq i}S_{j\varepsilon} G_0(x_i,x_j) +
    D|\Omega|^{-1}\sum_{j=1}^N S_{j\varepsilon}\,, & D\gg 1\,. \end{cases}
\end{equation}
>From the $D\gg 1$ case we see that $D={\mathcal O}(\varepsilon^{-1})$
is a distinguished regime for which the right-hand side has an
${\mathcal O}(1)$ contribution. Defining the vectors
$\pmb{S}_\varepsilon \equiv
(S_{1\varepsilon},\ldots,S_{N\varepsilon})^T$,
$\mu(\pmb{S}_\varepsilon) \equiv
(\mu(S_{1\varepsilon}),\ldots,\mu(S_{N\varepsilon}))^T$, and
$\pmb{e} \equiv (1,\ldots,1)^T$, as well as the matrices
$\mathcal{E}_N$, $\mathcal{G}$, and $\mathcal{G}_0$ by
\begin{equation}\label{eq:matrix_definitions}
  \mathcal{E}_N \equiv \frac{1}{N}\pmb{e}\pmb{e}^T\,,\qquad
  (\mathcal{G})_{ij} = \begin{cases} R(x_i)\,, & i=j\\ G(x_i,x_j)\,,& i\neq j
  \end{cases}\,,\qquad (\mathcal{G}_0)_{ij} = \begin{cases} R_0(x_i)\,,
    & i=j\\ G_0(x_i,x_j)\,,& i\neq j\end{cases}\,,
\end{equation}
we obtain from \eqref{eq:mu_all} that the unknowns
$S_{1\varepsilon},\ldots,S_{N\varepsilon}$ must satisfy the NAS
\begin{subequations}\label{eq:NAS}
\begin{align}
  \mu(\pmb{S}_\varepsilon) &=  4\pi\varepsilon\mathcal{G} \pmb{S}_\varepsilon\,,
  \qquad \mbox{for} \quad D={\mathcal O}(1)\,, \label{eq:NAS_order_1} \\
   \mu(\pmb{S}_\varepsilon) &= \kappa \mathcal{E}_N \pmb{S}_\varepsilon +
           4\pi\varepsilon \mathcal{G}_0\pmb{S}_\varepsilon \,, \qquad
    \mbox{for} \quad D=\varepsilon^{-1}D_0 \,, \qquad
    \mbox{where}  \quad \kappa \equiv \frac{4\pi N D_0}{|\Omega|}\,.
       \label{eq:NAS_order_1_over_epsilon}
\end{align}
\end{subequations}

\subsection{Symmetric and Asymmetric $N$-Spot Quasi-Equilibrium}

We now determine solutions to the NAS \eqref{eq:NAS} in
both the $D={\mathcal O}(1)$ and the
$D={\mathcal O}(\varepsilon^{-1})$ regimes. In particular, we show
that it is possible to construct \textit{symmetric} $N$-spot
solutions to \eqref{eq:pde_gm_3d} by finding a solution to the NAS
\eqref{eq:NAS} with $\pmb{S}_\varepsilon = S_{c\varepsilon}\pmb{e}$ in
both the $D={\mathcal O}(1)$ and $D={\mathcal O}(\varepsilon^{-1})$
regimes. Moreover, when $D={\mathcal O}(\varepsilon^{-1})$ we will
show that it is possible to construct \textit{asymmetric} quasi-equilibria
to \eqref{eq:pde_gm_3d} characterized by spots each having
one of two strengths.

When $D={\mathcal O}(1)$ the NAS \eqref{eq:NAS_order_1} implies that
to leading order $\mu(S_{i\varepsilon}) = 0$ for all
$i=1,\ldots,N$. From the properties of $\mu(S)$ outlined in
\S\ref{subsec:core_problem} and in particular the plot of $\mu(S)$ in
Figure \ref{fig:core_sol_mu}, we deduce that
$S_{i\varepsilon} \sim S_\star$ for all $i=1,\ldots,N$. Thus, to
leading order, $N$-spot quasi-equilibria in the $D={\mathcal O}(1)$
regime have spots with a common height, which we refer to as a \textit{
  symmetric} pattern. By calculating the next order term using
\eqref{eq:NAS_order_1} we readily obtain the two term result
\begin{equation}\label{eq:symm_sol_D_O_1}
  \pmb{S}_\varepsilon \sim S_\star \pmb{e} + \frac{4\pi \varepsilon S_\star}
  {\mu^{\prime}(S_\star)}\mathcal{G}\pmb{e}\,.
\end{equation}
We conclude that the configuration $x_1,\ldots,x_N$ of spots only
affects the spot strengths at ${\mathcal O}(\varepsilon)$ through the
Green's matrix $\mathcal{G}$. Note that if $\pmb{e}$ is an eigenvector
of $\mathcal{G}$ with eigenvalue $g_0$ then the solution to
\eqref{eq:NAS_order_1} is
$\pmb{S}_{i\varepsilon} = S_{c\varepsilon}\pmb{e}$ where
$S_{c\varepsilon}$ satisfies the scalar equation
$\mu(S_{c\varepsilon}) = 4\pi\varepsilon g_0 S_{c\varepsilon}$.

\begin{figure}[t!]
  \centering
	\begin{subfigure}[b]{0.33\textwidth}
		\centering
		\includegraphics[width=\textwidth,height=4.2cm]{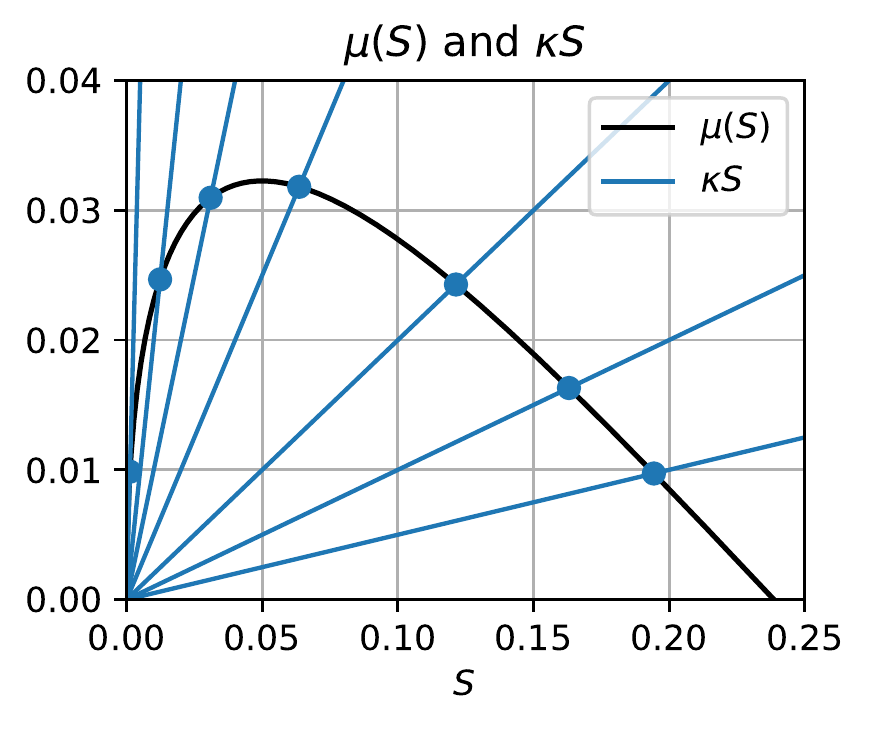}
		\caption{}
                \label{fig:symmetric_graphic_intersection}
	\end{subfigure}%
	\begin{subfigure}[b]{0.33\textwidth}
		\centering
		\includegraphics[width=\textwidth,height=4.2cm]{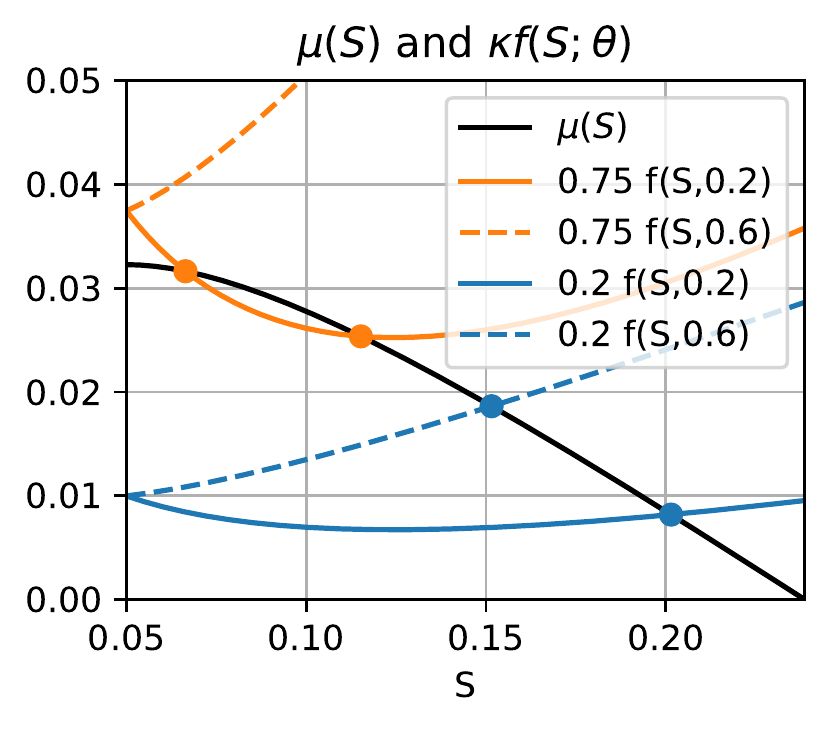}
		 \caption{}
                \label{fig:asymmetric_graphic_intersection}
	\end{subfigure}%
	\begin{subfigure}[b]{0.33\textwidth}
		\centering
		\includegraphics[width=\textwidth,height=4.2cm]{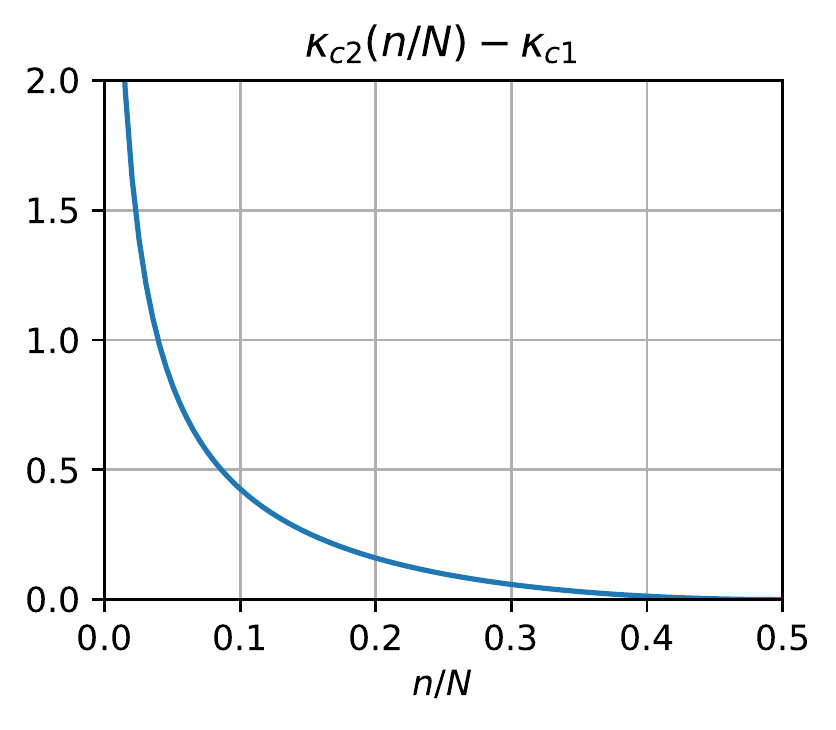}
		 \caption{}
                \label{fig:kappa_c2}
	\end{subfigure}%
	\caption{(a) Illustration of solutions to
          \eqref{eq:Sc_equation} as the intersection between $\mu(S)$
          and $\kappa S$. There is a unique solution if
          $\kappa < \kappa_{c1}\equiv {\mu(S_\text{crit}) /
            S_\text{crit}}$.  (b) Illustration of solutions to
          \eqref{eq:asymmetric_equation} as the intersection between
          $\mu(S)$ and $\kappa f(S,\theta)$ where $\theta=n/N$ denotes
          the fraction of \textit{large} spots in an asymmetric
          pattern. Note that when $\theta=0.2<0.5$ and
          $\kappa > \kappa_{c1}\approx 0.64619$ there exist two
          solutions. (c) Plot of $\kappa_{c2}-\kappa_{c1}$ versus
          $n/N$. Observe that $\kappa_{c2}-\kappa_{c1}$ increases as
          the fraction of large spots decreases.}
          \label{fig:graphic_illustration_solutions}
\end{figure}

\begin{figure}[t!]
  \centering
	\begin{subfigure}[b]{0.33\textwidth}
		\centering
		\includegraphics[width=\textwidth,height=4.2cm]{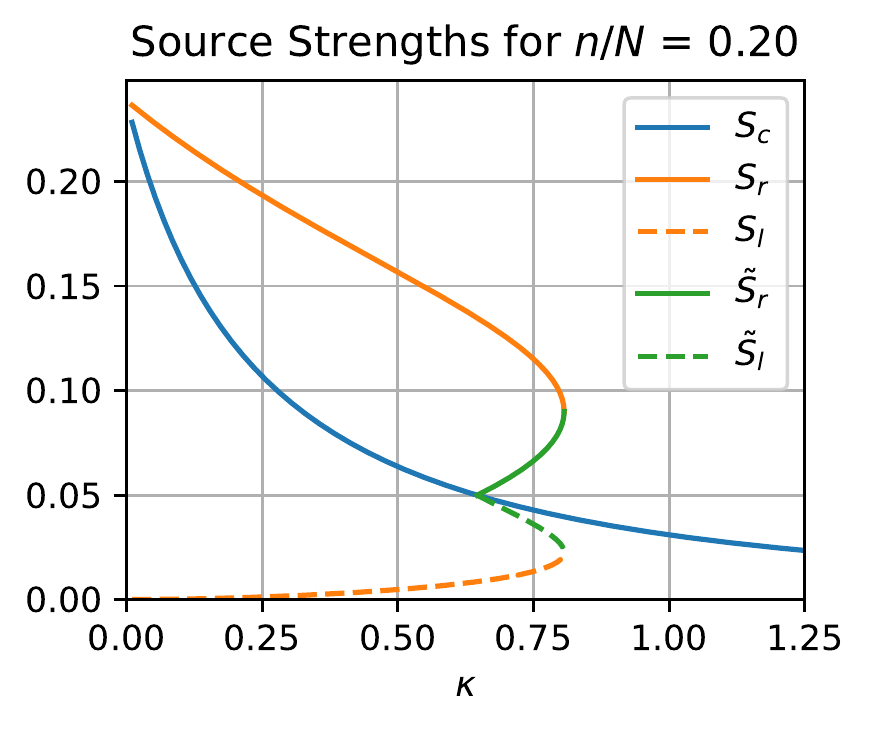}
		 \caption{}
                \label{fig:quasi_theta_20}
	\end{subfigure}%
	\begin{subfigure}[b]{0.33\textwidth}
		\centering
		\includegraphics[width=\textwidth,height=4.2cm]{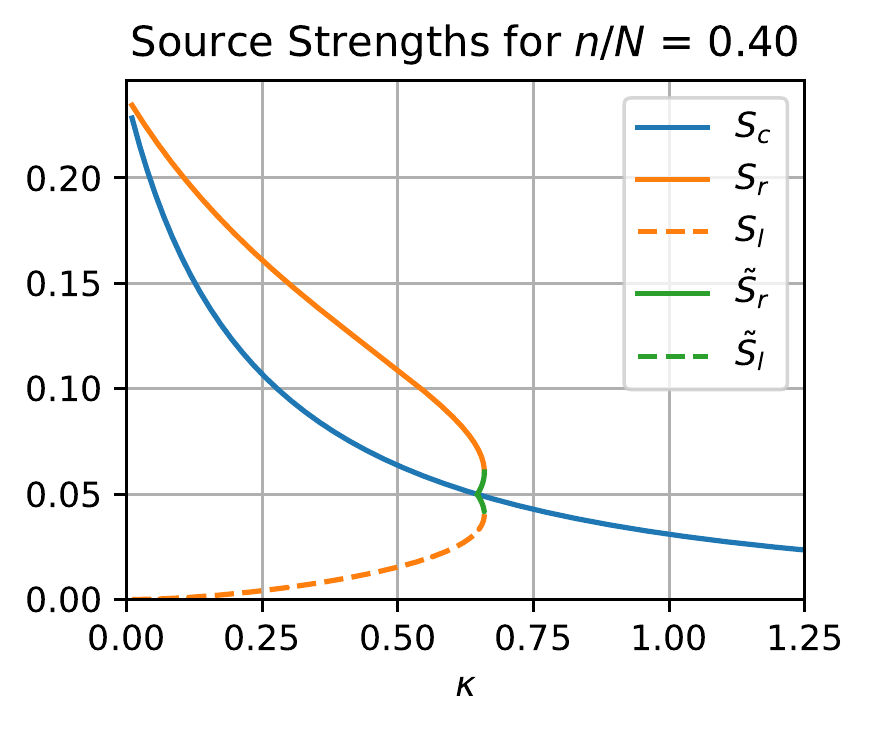}
		 \caption{}
                \label{fig:quasi_theta_40}
	\end{subfigure}%
	\begin{subfigure}[b]{0.33\textwidth}
		\centering
		\includegraphics[width=\textwidth,height=4.2cm]{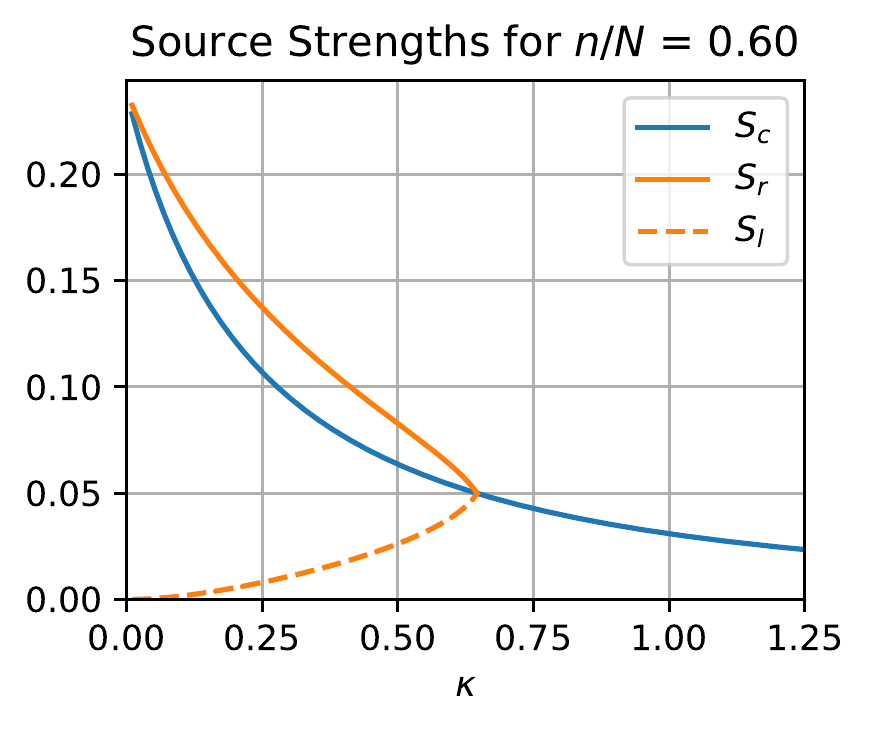}
		 \caption{}
                \label{fig:quasi_theta_60}
	\end{subfigure}%
	\caption{Bifurcation diagram illustrating the dependence on
          $\kappa$ of the common spot strength $S_c$ as well as the
          asymmetric spot strengths $S_r$ and $S_l$ or $\tilde{S}_r$
          and $\tilde{S}_l$. In (a) and (b) we have $n/N<0.5$ so that
          there are more small spots than large spots in
          an asymmetric pattern. As a result, we observe that there
          can be two types of asymmetric patterns with strengths $S_r$
          and $S_l$ or $\tilde{S}_r$ and $\tilde{S}_l$. In (c) the
          number of large spots exceeds that of small spots and only
          one type of asymmetric pattern is
          possible.}\label{fig:quasi_theta}
\end{figure}

Next, we consider solutions to the NAS
\eqref{eq:NAS_order_1_over_epsilon} in the $D=\varepsilon^{-1}D_0$
regime. Seeking a solution
$\pmb{S}_{\varepsilon} \sim \pmb{S}_{0} + \varepsilon\pmb{S}_{1} +
\cdots$ we obtain the leading order problem
\begin{equation}\label{eq:leading_order_NAS_large_D}
\mu(\pmb{S}_{0}) = \kappa \mathcal{E}_N\pmb{S}_{0}.
\end{equation}
Note that the concavity of $\mu(S)$ (see Figure \ref{fig:core_sol_mu})
implies the existence of two values $0<S_l<S_r<S_\star$ such that
$\mu(S_l)=\mu(S_r)$. Thus, in addition to the symmetric solutions
already encountered in the $D={\mathcal O}(1)$ regime, we also have
the possibility of \textit{asymmetric} solutions, where the spots can
have two different heights. We first consider symmetric
solutions, where to leading order $\pmb{S}_{0} = S_{c}\pmb{e}$ in which
$S_c$ satisfies 
\begin{equation}\label{eq:Sc_equation}
\mu(S_{c}) = \kappa S_c\,.
\end{equation}
The plot of $\mu(S)$ in Figure \ref{fig:core_sol_mu}, together with the
$S\ll 1$ asymptotics given in \eqref{eq:core_sol_small_S}, imply that a
solution to \eqref{eq:Sc_equation} can be found in the interval
$0<S_c\leq S_\star$ for all $\kappa>0$. In Figure
\ref{fig:symmetric_graphic_intersection} we illustrate graphically that
the common spot strength $S_c$ is obtained by the intersection of
$\mu(S)$ with the line $\kappa S$. We refer to Figure
\ref{fig:quasi_theta} for plots of the symmetric solution strengths as
a function of $\kappa$. In addition, we readily calculate that
\begin{equation}\label{eq:Sc_asymptotics}
  S_c \sim S_\star\biggl(1 + \frac{\kappa}{\mu^{\prime}(S_\star)}\biggr) +
  {\mathcal O}(\kappa^2)\,,\quad \mbox{for} \,\,\,  \kappa \ll 1\,;\qquad
  S_c \sim \frac{1}{b\kappa^2} + {\mathcal O}(\kappa^{-3})\,,\quad
  \mbox{for} \,\,\, \kappa \gg 1\,,
\end{equation}
which provide a connection between the $D={\mathcal O}(1)$ and
$D\rightarrow\infty$ (shadow limit) regimes, respectively. From
\eqref{eq:NAS_order_1_over_epsilon}, the next order correction $\pmb{S}_1$
satisfies $\mu^{\prime}(S_c) \pmb{S}_1 - \kappa \mathcal{E}_N \pmb{S}_1 =
4\pi S_c \mathcal{G}_0 \pmb{e}$. Upon left-multiplying this expression
by $\pmb{e}^T$ we can determine $\pmb{e}^T\pmb{S}_1$. Then, by recalling
the definition of $\mathcal{E}_N \equiv N^{-1}\pmb{e}\pmb{e}^T$ we can calculate
$\pmb{S}_1$. Summarizing, a two term asymptotic expansion for the
symmetric solution to \eqref{eq:NAS_order_1_over_epsilon} is 
\begin{equation}\label{eq:symm_sol_D_large}
  \pmb{S}_\varepsilon \sim S_c\pmb{e} + \frac{4\pi\varepsilon}{\mu^{\prime}(S_c)}
  \biggl( S_c \mathcal{I}_N + \frac{ \mu(S_c)}{\mu^{\prime}(S_c)-\kappa}
  \mathcal{E}_N \biggr) \mathcal{G}_0\pmb{e}\,,
\end{equation}
provided that $\mu^{\prime}(S_c)\neq 0$ (i.e.
$S_c\neq S_\text{crit}$). Note that $\mu^{\prime}(S_c) - \kappa = 0$
is impossible by the following simple argument. First, for this
equality to hold we require that $0<S<S_\text{crit}$ since otherwise
$\mu^{\prime}(S_c)<0$. Moreover, we can solve \eqref{eq:Sc_equation}
for $\kappa$ to get $\mu^{\prime}(S_c)-\kappa = S_c^{-1}g(S_c)$ where
$g(S) \equiv S\mu^{\prime}(S) - \mu(S)$. However, we calculate
$g^{\prime}(S) = S\mu^{\prime\prime}(S) < 0$ and moreover, using the
small $S$ asymptotics found in \eqref{eq:core_sol_small_S} we
determine that $g(S) \sim -\sqrt{S/(4b)} < 0$ as $S\rightarrow
0^+$. Therefore, $g(S_c)<0$ for all $0<S_c<S_\text{crit}$ so that 
$\mu^{\prime}(S_c) < \kappa$ holds. Finally, as for the
$D={\mathcal O}(1)$ case, if $\mathcal{G}_0\pmb{e}=g_{00}\pmb{e}$
then the common source values extends to higher order and we have
$\pmb{S}_\varepsilon = S_{c\varepsilon} \pmb{e}$ where
$S_{c\varepsilon}$ is the unique solution to the scalar problem
\begin{equation}\label{eq:Sc_eps_equation}
\mu(S_{c\varepsilon}) = (\kappa  + 4\pi\varepsilon g_{00})S_{c\varepsilon}\,.
\end{equation}

\begin{figure}[t!]
  \centering
	\begin{subfigure}[b]{0.33\textwidth}
		\centering
		\includegraphics[width=\textwidth,height=4.2cm]{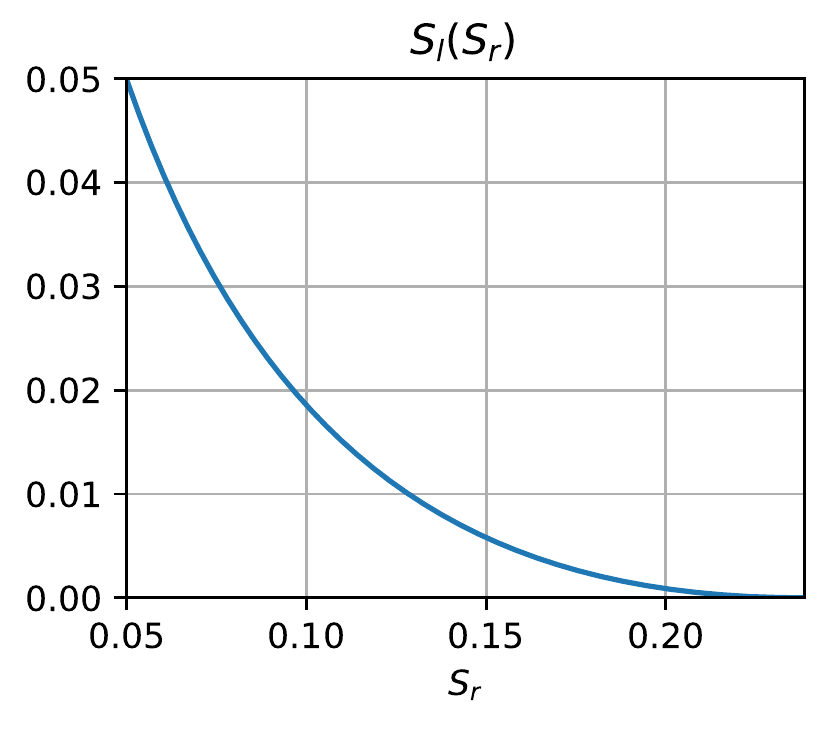}
		 \caption{}
                \label{fig:Sl}
	\end{subfigure}%
	\begin{subfigure}[b]{0.33\textwidth}
		\centering
		\includegraphics[width=\textwidth,height=4.2cm]{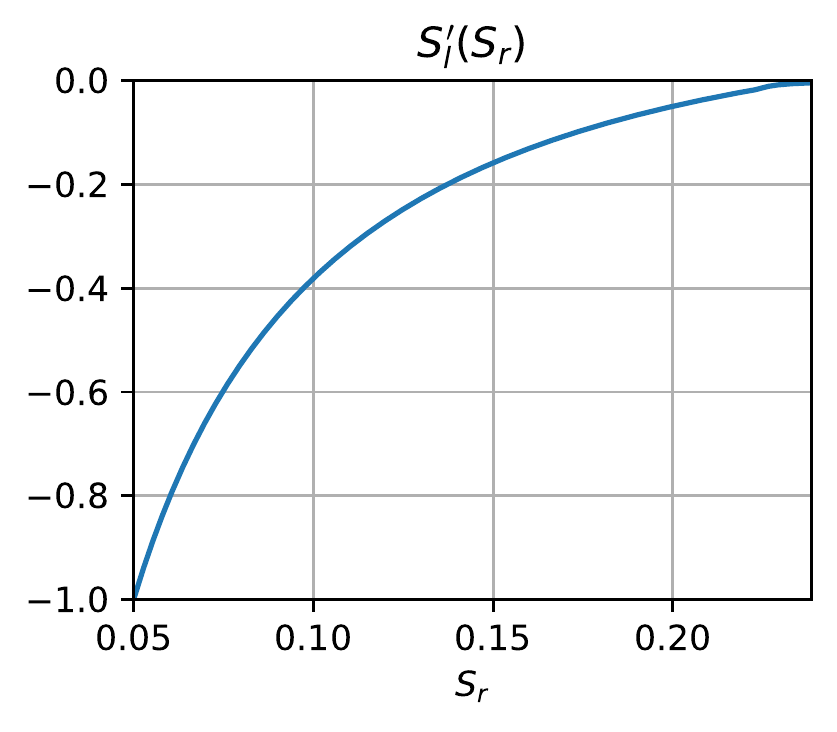}
	 \caption{}
                \label{fig:Sl_prime}
	\end{subfigure}%
	\begin{subfigure}[b]{0.33\textwidth}
		\centering
		\includegraphics[width=\textwidth,height=4.2cm]{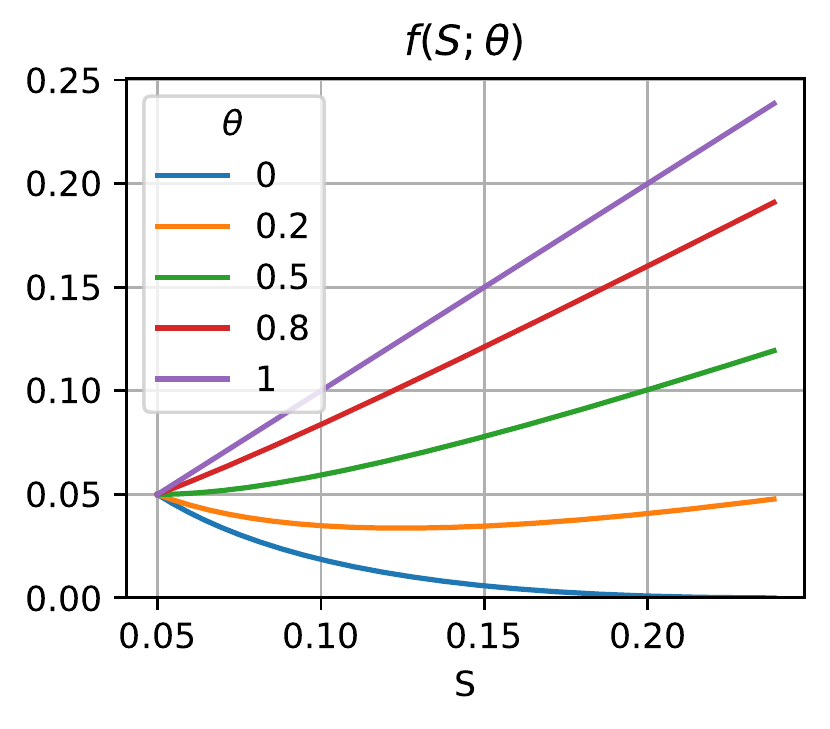}
		\caption{}
                \label{fig:f_s_theta}
	\end{subfigure}%
	\caption{Plots of (a) $S_l(S_r)$ and (b) $S_l^{\prime}(S_r)$
          for the construction of asymmetric $N$-spot patterns. (c)
          Plots of $f(S,\theta)$ for select values of
          $\theta\equiv {n/N}$. For $0<\theta<0.5$ the function
          $f(S,\theta)$ attains an interior minimum in
          $S_\text{crit}<S<S_\star$.}\label{fig:s_s_prime_f}
\end{figure}

Next, we construct of \textit{asymmetric} $N$-spot configurations. The
plot of $\mu(S)$ indicates that for any value of
$S_r\in(S_\text{crit},S_\star]$ there exists a unique value
$S_l = S_l(S_r) \in [0,S_\text{crit})$ satisfying
$\mu(S_l) = \mu(S_r)$. A plot of $S_l(S_r)$ is shown in Figure
\ref{fig:Sl}. Clearly $S_l(S_\text{crit}) = S_\text{crit}$ and
$S_l(S_\star) = 0$. We suppose that to leading order the $N$-spot
configuration has $n$ \textit{large} spots of strength $S_r$ and $N-n$
\textit{small} spots of strengths $S_l$. More specifically, we seek a
solution of the form
\begin{equation}\label{eq:asymmetric_sol_1}
\pmb{S}_\varepsilon \sim (S_r,\ldots,S_r,S_l(S_r),\ldots,S_l(S_r))^T \,,
\end{equation}
so that \eqref{eq:leading_order_NAS_large_D} reduces to the single
scalar nonlinear equation
\begin{equation}\label{eq:asymmetric_equation}
  \mu(S_r) = \kappa f(S_r;n/N) \,,  \quad \mbox{on} \quad
  S_\text{crit}<S_r<S_\star \,, \qquad \mbox{where}
   \qquad f(S;\theta) \equiv \theta S + (1-\theta) S_l(S)\,.
\end{equation}
Since
$\mu(S_\text{crit}) - \kappa f(S_\text{crit};n/N) = \mu(S_\text{crit})
- \kappa S_\text{crit}$ and
$\mu(S_\star) - \kappa f(S_\star;n/N) = -\kappa {nS_\star/N}<0$, we
obtain by the intermediate value theorem that there exists
at least one solution to \eqref{eq:asymmetric_equation} for any
$0<n\leq N$ when
$$
0 < \kappa < \kappa_{c1}\equiv {\mu(S_\text{crit}) / S_\text{crit}}
\approx 0.64619 \,.
$$
Next, we calculate
$$
f^{\prime}(S;\theta) = (1-\theta)\biggl( \frac{\theta}{1-\theta} +
S_l^{\prime}(S)\biggr)\,,
$$
where $S_l^{\prime}(S)$ is computed numerically (see Figure
\ref{fig:Sl_prime}). We observe that $-1\leq S_l^{\prime}(S_r)\leq 0$
with $S_l^{\prime}(S_\text{crit}) = -1$ and
$S_l^{\prime}(S_\star) = 0$. In particular, $f(S;n/N)$ is monotone
increasing if $\theta/(1-\theta)={n/(N-n)} > 1$, while it attains a
local minimum in $(S_\text{crit},S_\star)$ if $n/(N-n) < 1$. A plot of
$f(S;\theta)$ is shown in Figure \ref{fig:f_s_theta}. In either case, we
deduce that the solution to \eqref{eq:asymmetric_equation} when
$0<\kappa<\kappa_{c1}$ is unique (see
Figure \ref{fig:symmetric_graphic_intersection}). On the other hand,
when $n/(N-n)<1$ we anticipate an additional range of values
$\kappa_{c1} < \kappa < \kappa_{c2}$ for which
\eqref{eq:asymmetric_equation} has \textit{two} distinct solutions
$S_\text{crit}<\tilde{S}_r<S_r<S_\star$. Indeed, this threshold can be
found by demanding that $\mu(S)$ and $\kappa f(S;n/N)$ intersect
tangentially. In this way, we find that the threshold
$\kappa_{c2}$ can be written as
\begin{subequations}\label{eq:kappa_c2_system}
\begin{equation}
  \kappa_{c2}= \kappa_{c2}(n/N)\equiv \frac{\mu(S_r^\star)}{f(S_r^\star;n/N)}\,,
\end{equation}
where $S_r^{\star}$ is the unique solution to
\begin{equation}
f(S_r^\star;n/N)\mu^{\prime}(S_r^\star) = f^{\prime}(S_r^\star;n/N)\mu(S_r^\star) \,.
\end{equation}
\end{subequations}

In Figure \ref{fig:kappa_c2} we plot $\kappa_{c2}-\kappa_{c1}$ as a
functions of $n/N$ where we observe that $\kappa_{c2}>\kappa_{c1}$
with $\kappa_{c2}-\kappa_{c1}\rightarrow 0^+$ and
$\kappa_{c2}-\kappa_{c1}\rightarrow \infty$ as $n/N\rightarrow 0.5^-$
and $n/N\rightarrow 0^+$ respectively. Furthermore, in Figure
\ref{fig:asymmetric_graphic_intersection} we graphically illustrate
how multiple solutions to \eqref{eq:asymmetric_equation} arise as
$\theta=n/N$ and $\kappa$ are varied. We remark that the condition
$n/(N-n) < 1$ implies that $n<{N/2}$, so that there are more small
than large spots. The appearance of two distinct
asymmetric patterns in this regime has a direct analogy to results
obtained for the 1-D and 2-D GM model in \cite{ward_2002_asymmetric}
and \cite{WGM1}, respectively. The resulting bifurcation diagrams are
shown in Figure \ref{fig:quasi_theta} for $n/N = 0.2, 0.4, 0.6$. We
summarize our results for quasi-equilibria in the following
proposition.

\begin{proposition}\label{prop:quasiequilibrium}(Quasi-Equilibria):
  Let $\varepsilon\rightarrow 0$ and $x_1,\ldots,x_N\in\Omega$ be
  well-separated. Then, the 3-D GM model \eqref{eq:pde_gm_3d} admits an
  $N$-spot quasi-equilibrium solution with inner asymptotics
\begin{equation}\label{eq:inner_asym}
v\sim D V_i(\varepsilon^{-1}|x-x_i|)\,,\qquad
u\sim D U_i(\varepsilon^{-1}|x-x_i|)\,,
\end{equation}
as $x\rightarrow x_i$ for each $i=1,\ldots,N$ where $V_i$ and $U_i$
are given by \eqref{eq:sol_inner}. When $|x-x_i|={\mathcal O}(1)$, the
activator is exponentially small while the inhibitor is given by
\eqref{eq:u_outer}. The spot strengths $S_{i\varepsilon}$ for
$i=1,\ldots,N$ completely determine the asymptotic solution and there
are two distinguished limits. When $D={\mathcal O}(1)$ the spot
strengths satisfy the NAS \eqref{eq:NAS_order_1}, which has the
leading order asymptotics \eqref{eq:symm_sol_D_O_1}. In
particular, $S_{i\varepsilon}\sim S_{\star}$ so all $N$-spot patterns
are symmetric to leading order. When $D=\varepsilon^{-1}D_0$ the spot
strengths satisfy the NAS \eqref{eq:NAS_order_1_over_epsilon}. A
symmetric solution with asymptotics \eqref{eq:symm_sol_D_large} where
$S_c$ satisfies \eqref{eq:Sc_equation} always exists. Moreover, if
$$
0 < \frac{4\pi N D_0}{|\Omega|} < \kappa_{c1} \approx 0.64619\,,
$$
then an asymmetric pattern with $n$ large spots of strength
$S_r\in(S_\text{crit},S_\star)$ and $N-n$ small spots of strength
$S_l\in(0,S_\text{crit})$ can be found by solving
\eqref{eq:asymmetric_equation} for $S_r$ and calculating $S_l$ from
$\mu(S_l)=\mu(S_r)$. If, in addition we have $n/(N-n)<1$, then
\eqref{eq:asymmetric_equation} admits two solutions on the range
$$
0.64619 \approx \kappa_{c1} < \frac{4\pi N D_0}{|\Omega|} < \kappa_{c2}(n/N)\,,
$$
where $\kappa_{c2}(n/N)$ is found by solving the system
\eqref{eq:kappa_c2_system}.
\end{proposition}

As we have already remarked, in the $D=D_0/\varepsilon$ regime, if
$D_0\ll 1$ then the symmetric $N$-spot solution
\eqref{eq:symm_sol_D_large} coincides with the symmetric solution for
the $D={\mathcal O}(1)$ regime given by \eqref{eq:symm_sol_D_O_1}.
The asymmetric solutions predicted for the $D=D_0/\varepsilon$ regime
persist as $D_0$ decreases and it is, therefore, natural to ask what
these solutions correspond to in the $D={\mathcal O}(1)$ regime. From
the small $S$ asymptotics \eqref{eq:core_sol_small_S} we note that the
NAS \eqref{eq:NAS_order_1} does admit an asymmetric solution, albeit
one in which the source strengths of the small spots are of
${\mathcal O}(\varepsilon^2)$.  Specifically, for a given integer
$n$ in $1<n\leq N$ we can construct a solution where
\begin{equation}
  \pmb{S}_\varepsilon \sim (S_\star,\ldots,S_\star,\varepsilon^2 S_{n+1,0},\ldots,
  \varepsilon^2 S_{N,0})^T \,.
\end{equation}
By using the small $S$ asymptotic expansion for $\mu(S)$ given in
\eqref{eq:core_sol_small_S}, we obtain from \eqref{eq:NAS_order_1} that
\begin{equation}\label{eq:Si0_temp}
  S_{i,0} = b \left(4\pi S_\star \sum_{j=1}^n G(x_i,x_j)\right)^2\,,\qquad
  i=n+1,\ldots,N \,.
\end{equation}
We observe that in order to support $N-n$ spots of strength
${\mathcal O}(\varepsilon^2)$, we require at least one spot
of strength ${\mathcal O}(1)$. Setting $D = {D_0/\varepsilon}$, we
use the large $D$ asymptotics for $G(x,\xi)$ in
\eqref{eq:greens_function_large_D} to reduce \eqref{eq:Si0_temp} to
\begin{equation}
  S_{i,0} \sim b\varepsilon^{-2}\biggl(\frac{4\pi D_0 n S_\star}{|\Omega|}\biggr)^2
  \,, \qquad i=n+1,\ldots,N \,.
\end{equation}
Alternatively, by taking $\kappa \ll 1$ in the NAS
\eqref{eq:NAS_order_1_over_epsilon} for the $D={D_0/\varepsilon}$
regime, we conclude that $S_r \sim S_\star$ and
$S_l \sim b\left({\kappa n S_\star/N}\right)^2$.  Since
$\kappa n/N = 4\pi D_0 n/|\Omega|$, as obtained from
\eqref{eq:NAS_order_1_over_epsilon}, we confirm that the asymmetric
patterns in the $D={D_0/\varepsilon}$ regime lead to an asymmetric
pattern consisting of spots of strength ${\mathcal O}(1)$ and
${\mathcal O}(\varepsilon^2)$ in the $D={\mathcal O}(1)$ regime.

\section{Linear Stability}\label{sec:stability}

Let $(v_{qe},u_{qe})$ be an $N$-spot quasi-equilibrium solution as
constructed in \S \ref{sec:quasi}. We will analyze instabilities for
quasi-equilibria that occur on ${\mathcal O}(1)$ time-scales. To do
so, we substitute
\begin{equation}
v = v_{qe} + e^{\lambda t}\phi\,,\qquad u = u_{qe} + e^{\lambda t}\psi\,,
\end{equation}
into \eqref{eq:pde_gm_3d} and, upon linearizing, we obtain the
eigenvalue problem
\begin{equation}\label{lin:eig_all}
  \varepsilon^2 \Delta \phi - \phi + \frac{2 v_{qe}}{u_{qe}}\phi -
  \frac{v_{qe}^2}{u_{qe}^2}\psi = \lambda \phi\,,\qquad D\Delta\psi -
  \psi + 2\varepsilon^{-2} v_{qe} \phi = \tau\lambda\psi\,,
\end{equation}
where $\partial_n\phi=\partial_n\psi=0$ on $\partial\Omega$. In the
inner region near the $j^\text{th}$ spot, we introduce a local
expansion in terms of the associated Legendre polynomials
$P_l^m(\cos\theta)$ of degree $l=0,2,3,\dots,$ and order
  $m=0,1,\dots,l$
\begin{equation}\label{eq:inner_eig}
  \phi \sim c_j  D P_l^m(\cos\theta)e^{i m\varphi} \Phi_j(\rho)\,,\qquad
  \psi \sim c_j D P_l^m(\cos\theta) e^{im\varphi}\Psi_j(\rho)\,,
\end{equation}
where $\rho = \varepsilon^{-1}|x-x_j|$, and
$(\theta,\varphi)\in (0,\pi)\times[0,2\pi)$. Suppressing subscripts
for the moment, and assuming that $\varepsilon^2\tau\lambda/D \ll 1$,
we obtain the leading order inner problem
\begin{subequations}\label{lin:psi_inner}
\begin{equation}\label{eq:linear_stability_pde}
  \Delta_\rho\Phi - \frac{l(l+1)}{\rho^2}\Phi - \Phi + \frac{2 V}{U}\Phi -
  \frac{V^2}{U^2}\Psi = \lambda \Phi\,,\qquad
  \Delta_\rho\Psi - \frac{l(l+1)}{\rho^2}\Psi + 2V\Phi = 0\,, \quad \rho>0 \,,
\end{equation}
with the boundary conditions $\Phi^{\prime}(0)=\Psi^{\prime}(0)=0$,
and $\Phi\rightarrow 0$ as $\rho\rightarrow\infty$. Here $(V,U)$
satisfy the core problem \eqref{eq:core}. The behaviour of $\Psi$ as
$\rho\rightarrow\infty$ depends on the parameter $l$. More
specifically, we have that
\begin{equation}\label{eq:decay}
  \Psi \sim \begin{cases} B(\lambda,S) + \rho^{-1}\,, & \,\,\mbox{for}
    \,\,\,\, l = 0\,, \\
    \rho^{-(1/2 + \gamma_l)}\,, & \,\,\mbox{for}\,\,\,\,  l > 0\,,
  \end{cases}\qquad \mbox{as} \quad\rho\rightarrow\infty\,,
\end{equation}
\end{subequations}
where $\gamma_l \equiv \sqrt{\tfrac{1}{4}+l(l+1)}$ and
$B(\lambda,S)$ is a constant. Here we have normalized $\Psi$ by fixing
to unity the multiplicative factor in the decay rate in \eqref{eq:decay}.
Next, we introduce the Green's function $G_l(\rho,\tilde{\rho})$ solving
\begin{equation}\label{eq:green_l}
  \Delta_\rho G_l - \frac{l(l+1)}{\rho^2} G_l = -
  \rho^{-2}\delta(\rho-\tilde{\rho}) \,, \quad \mbox{given by} \quad
  G_l(\rho,\tilde{\rho}) = \frac{1}{2\gamma_l\sqrt{\rho\tilde{\rho}}}
  \begin{cases}
    (\rho/\tilde{\rho})^{\gamma_l}\,, & 0<\rho<\tilde{\rho}\,, \\
    (\tilde{\rho}/\rho)^{\gamma_l}\,, & \rho > \tilde{\rho}\,,\end{cases}
\end{equation}
when $l>0$. For $l=0$ the same expression applies, but an arbitrary
constant may be added. For convenience we fix this constant to be
zero. In terms of this Green's function we can solve for $\Psi$ 
explicitly in \eqref{eq:linear_stability_pde} as
\begin{equation}\label{eq:Psi_solution_far}
  \Psi = 2\int_0^\infty G_l(\rho,\tilde{\rho}) V(\tilde{\rho})
  \Phi(\tilde{\rho}) \tilde{\rho}^2 \, d\tilde{\rho} +
  \begin{cases}
    B(\lambda,S)\,, & \,\, \mbox{for} \,\,\,\, l=0\,,\\
    0\,, & \,\, \mbox{for} \,\,\,\, l>0\,.
  \end{cases}
\end{equation}
Upon substituting this expression into \eqref{eq:linear_stability_pde} we
obtain the nonlocal spectral problems
\begin{subequations}
\begin{equation}\label{eq:M0_problem}
  \mathscr{M}_0\Phi = \lambda \Phi + B(\lambda,S)\frac{V^2}{U^2} \,, \quad
  \mbox{for}\quad l=0 \,; \qquad \mathscr{M}_l\Phi = \lambda \Phi\,, \quad
  \mbox{for}\quad l>0 \,.
\end{equation}
Here the integro-differential operator $\mathscr{M}_l$ is defined for every
$l\geq 0$ by
\begin{equation}\label{eq:ml_operator}
  \mathscr{M}_l\Phi \equiv \Delta_\rho \Phi -\frac{l(l+1)}{\rho^2}\Phi - \Phi
  + \frac{2V}{U}\Phi - \frac{2V^2}{U^2}\int_0^\infty
  G_l(\rho,\tilde{\rho})V(\tilde{\rho})\Phi(\tilde{\rho})\tilde{\rho}^2 \,
  d\tilde{\rho}\,.
\end{equation}
\end{subequations}

\begin{figure}[t!]
	\begin{subfigure}[b]{0.49\textwidth}
		\centering
		\includegraphics[width=\textwidth,height=4.2cm]{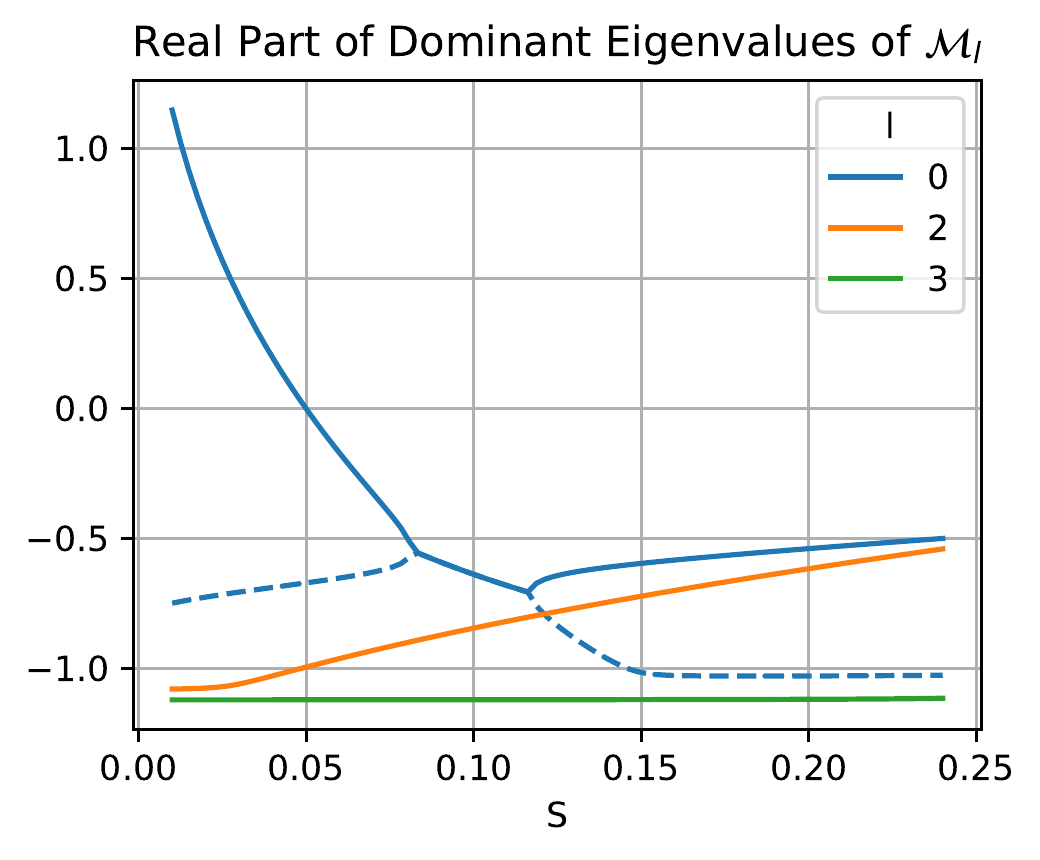}
		 \caption{}
                \label{fig:spectrum_Ml}
	\end{subfigure}%
	\begin{subfigure}[b]{0.49\textwidth}
		\centering
		\includegraphics[width=\textwidth,height=4.2cm]{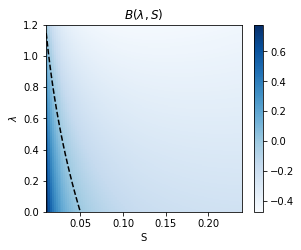}
                \caption{}
                \label{fig:B_continuous}
	\end{subfigure}%
	\caption{(a) Spectrum of the operator $\mathscr{M}_l$ defined
          in \eqref{eq:ml_operator}. The dashed blue line indicates
          the eigenvalue with second largest real part for
          $l=0$. Notice that the dominant eigenvalue of
          $\mathscr{M}_0$ is zero when
          $S=S_\text{crit}\approx 0.04993$, corresponding to the
          maximum of $\mu(S)$ (see Figure \ref{fig:core_sol_mu}).  (b)
          Plot of $B(\lambda,S)$. The dashed line black indicates the
          largest positive eigenvalue of $\mathscr{M}_0(S)$ and also
          corresponds to the contour $B(\lambda,S) = 0$. We observe that
          $B(\lambda,S)$ is both continuous and negative for
          $S>S_\text{crit} \approx 0.04993$.}\label{fig:prop_B}
\end{figure}

A key difference between the $l=0$ and $l>0$ linear stability problems
is the appearance of an unknown constant $B(\lambda,S)$ in the $l=0$
equation. This unknown constant is determined by matching the
far-field behaviour of the inner inhibitor expansion with the outer
solution. In this sense, we expect that $B(\lambda,S)$ will
encapsulate global contributions from all spots, so that instabilities
for the mode $l=0$ are due to the interactions between spots. In
contrast, the absence of an unknown constant for instabilities for the
$l>0$ modes indicates that these instabilities are localized, and that
the weak effect of any interactions between spots occurs only through
higher order terms. In this way, instabilities for modes with $l>0$
are determined solely by the spectrum of the operator
$\mathscr{M}_l$. In Figure \ref{fig:spectrum_Ml} we plot the
numerically-computed dominant eigenvalue of $\mathscr{M}_l$ for
$l=0,2,3$ as well as the sub dominant eigenvalue for $l=0$ for
$0<S<S_\star$. This spectrum is calculated from the discretization of
$\mathscr{M}_l$ obtained by truncating the infinite domain to
$0<\rho<L$, with $L\gg 1$, and using a finite difference approximation
for spatial derivatives combined with a trapezoidal rule
discretization of the integral terms. The $l=1$ mode always admits a
zero eigenvalue, as this simply reflects the translation invariance of
the inner problem. Indeed, these instabilities will be briefly
considered in Section \ref{sec:slow_dynamics} where we consider the
slow dynamics of quasi-equilibrium spot patterns. From Figure
\ref{fig:spectrum_Ml} we observe that the dominant eigenvalues of
$\mathscr{M}_l$ for $l=2,3$ satisfy $\mbox{Re}(\lambda)<0$
(numerically we observe the same for larger values of $l$). Therefore,
since the modes $l>1$ are always \textit{linearly stable}, for the 3-D
GM model there will be no \textit{peanut-splitting} or spot
self-replication instabilities such as observed for the 3-D
Schnakenberg model in \cite{tzou_2017_schnakenberg}. In the next
subsection we will focus on analyzing instabilities associated with
$l=0$ mode, which involves a global coupling between localized spots.

\subsection{Competition and Hopf Instabilities for the $l=0$ Mode}

>From \eqref{eq:M0_problem} we observe that $\lambda$ is in the
spectrum of $\mathscr{M}_0$ if and only if $B(\lambda,S)=0$. Assuming
that $B(\lambda,S)\neq 0$ we can then solve for $\Phi$ in
\eqref{eq:M0_problem} as
\begin{equation}\label{eq:Phi_sol}
\Phi = B(\lambda,S)(\mathscr{M}_0-\lambda)^{-1}(V^2/U^2) \,.
\end{equation}
Upon substituting \eqref{eq:Phi_sol} into the expression
\eqref{eq:Psi_solution_far} for $\Psi$ when $l=0$, we let $\rho\to\infty$
and use $G_0(\rho,\tilde{\rho}) \sim {1/\rho}$ as $\rho\to \infty$, as
obtained from \eqref{eq:green_l}, to deduce the far-field behavior
\begin{equation}\label{lin:psi_inf}
  \Psi \sim B + \frac{2B}{\rho}\int_0^\infty
  V (\mathscr{M}_0-\lambda)^{-1}(V^2/U^2)\rho^2d\,\rho\,, \qquad
  \mbox{as} \quad \rho\rightarrow\infty \,.
\end{equation}
We compare this expression with the normalized decay condition on
$\Psi$ in \eqref{eq:decay} for $l=0$ to conclude that
\begin{equation}\label{eq:B_equation}
  B(\lambda,S) = \frac{1}{2\int_0^\infty V (\mathscr{M}_0-\lambda)^{-1}(V^2/U^2)
    \rho^2\,d\rho} \,.
\end{equation}

We now solve the outer problem and through a matching condition derive
an algebraic equation for the eigenvalue $\lambda$. Since the
interaction of spots will be important for analyzing instabilities for
the $l=0$ mode, we re-introduce the subscript $j$ to label the spot.
First, since $\partial_\rho \Psi_j\sim -\rho^{-2}$ as
$\rho\to \infty$, as obtained from
\eqref{eq:decay} for $l=0$, an application of the divergence theorem
to $\Delta_\rho\Psi_j = - 2V_j\Phi_j$ yields that
$\int_0^\infty V_j\Phi_j\rho^2 \, d\rho = 1/2$. Next, by using
$v_{qe}\sim DV_j(\rho)$ and $\phi\sim c_jD\Phi_j(\rho)$ for
$|x-x_j|={\mathcal O}(\varepsilon)$ as obtained from
\eqref{eq:inner_asym} and \eqref{eq:inner_eig}, respectively, we
calculate in the sense of distributions for $\varepsilon\to 0$ that
\begin{equation*}
  2\varepsilon^{-2}v_{qe}\phi\rightarrow 8\pi \varepsilon D^2
  \sum_{j=1}^{N} c_j \left(\int_{0}^{\infty} V_j\Phi_j \rho^2\, d\rho \right)\,
  \delta(x-x_j) = 4\pi \varepsilon D^2 \sum_{j=1}^{N} c_j \delta(x-x_j) \,.
\end{equation*}
Therefore, by using this distributional limit in the equation for
$\psi$ in \eqref{lin:eig_all}, the outer problem for $\psi$ is
\begin{equation}\label{lin:psi_out}
  \Delta \psi - \frac{(1+\tau\lambda)}{D} \psi  =
  -4\pi\varepsilon D\sum_{j=1}^N c_j \delta(x-x_j)\,,
  \quad x\in\Omega\,; \qquad \partial_n\psi=0\,,\quad x\in\partial\Omega\,.
\end{equation}
The solution to \eqref{lin:psi_out} is represented as
\begin{equation}\label{lin:psi_outer}
\psi = 4\pi \varepsilon D \sum_{j=1}^N c_j G^\lambda(x,x_j) \,,
\end{equation}
where $G^\lambda(x,\xi)$ is the eigenvalue-dependent Green's function
satisfying
\begin{equation}\label{lin:green_lambda}
\begin{split}
  \Delta G^\lambda - \frac{(1+\tau\lambda)}{D} G^\lambda &= -\delta(x-\xi)\,, \quad
  x\in\Omega\,;\qquad \partial_n G^\lambda = 0\,,\quad x\in\partial\Omega\,,\\
  G^\lambda(x,\xi) &\sim \frac{1}{4\pi|x-\xi|} + R^\lambda(\xi) + o(1)\,,
  \qquad\mbox{as}\quad x \to \xi \,.
\end{split}
\end{equation}
By matching the limit as $x\rightarrow x_i$ of $\psi$ in
\eqref{lin:psi_outer} with the far-field behaviour
$\psi\sim D c_i B(\lambda,S_i)$ of the inner solution, as obtained
from \eqref{lin:psi_inf} and \eqref{eq:inner_eig}, we obtain the
matching condition
\begin{equation}\label{eq:B_matching}
  B(\lambda, S_i) c_i = 4\pi\varepsilon\biggl( c_i R^\lambda(x_i) +
  \sum_{j\neq i}^{N} c_j G^\lambda(x_i,x_j)\biggr)\,.
\end{equation}
As similar to the construction of quasi-equilibria in \S
\ref{sec:quasi}, there are two distinguished limits
$D={\mathcal O}(1)$ and $D={D_0/\varepsilon}$ to consider. The
stability properties are shown to be significantly different in
these two regimes.

In the $D={\mathcal O}(1)$ regime, we recall that $S_i\sim S_\star$
for $i=1,\ldots,N$ where $\mu(S_\star) = 0$. From
\eqref{eq:B_matching}, we conclude to leading order that
$B(\lambda,S_\star)=0$, so that $\lambda$ must be an eigenvalue of
$\mathscr{M}_0$ when $S=S_\star$. However, from Figure
\ref{fig:spectrum_Ml} we find that all eigenvalues of $\mathscr{M}_0$
when $S=S_\star$ satisfy $\mbox{Re}(\lambda)<0$. As such, from our
leading order calculation we conclude that $N$-spot quasi-equilibria
in the $D={\mathcal O}(1)$ regime are all linearly stable.

For the remainder of this section we focus exclusively on the
$D={D_0/\varepsilon}$ regime. Assuming that
$\varepsilon|1+\tau\lambda|/D_0 \ll 1$ we calculate
$G^\lambda(x,\xi) \sim {\varepsilon^{-1}
  D_0/\left[(1+\tau\lambda)|\Omega|\right]} + G_0(x,\xi)$, where $G_0$
is the Neumann Green's function satisfying \eqref{eq:G0_equation}. We
substitute this limiting behavior into \eqref{eq:B_matching} and, after
rewriting the the resulting homogeneous linear system for
$\pmb{c}\equiv (c_1,\ldots,c_N)^T$ in matrix form, we obtain 
\begin{equation}\label{eq:B_matching_D0}
  {\mathcal B} \pmb{c} = \frac{\kappa}{1+\tau\lambda}
  \mathcal{E}_N\pmb{c} + 4\pi\varepsilon\mathcal{G}_0 \pmb{c} \,, \qquad
  \mbox{where} \quad
  {\mathcal B}\equiv  \mbox{diag}(B(\lambda,S_1),\ldots,B(\lambda,S_N)) \,,
  \quad \mathcal{E}_N \equiv N^{-1}\pmb{e}\pmb{e}^T \,.
 \end{equation}
Here ${\mathcal G}_0$ is the Neumann Green's matrix and
$\kappa\equiv {4\pi ND_0/|\Omega|}$ (see
\eqref{eq:NAS_order_1_over_epsilon}).  Next, we separate the
proceeding analysis into the two cases: symmetric quasi-equilbrium
patterns and asymmetric quasi-equilibria.

\subsubsection{Stability of Symmetric Patterns in the $D=D_0/\varepsilon$ Regime}

We suppose that the quasi-equilibrium solution is symmetric so that to
leading order $S_1=\ldots =S_N=S_c$ where $S_c$ is found by solving
the nonlinear algebraic equation \eqref{eq:Sc_equation}. Then, from
\eqref{eq:B_matching_D0}, the leading order stability problem is
\begin{equation}\label{eq:B_matching_D0_symmetric}
B(\lambda,S_c)\pmb{c} = \frac{\kappa}{1+\tau\lambda}\mathcal{E}_N\pmb{c}\,.
\end{equation}

We first consider \textit{competition} instabilities for $N\geq 2$
characterized by $\pmb{c}^T \pmb{e} = 0$ so that
${\mathcal E}_N\pmb{c}=0$. Since $B(\lambda,S_c)=0$ from
\eqref{eq:B_matching_D0_symmetric}, it follows that $\lambda$ must be
an eigenvalue of $\mathscr{M}_0$, defined in \eqref{eq:ml_operator},
at $S=S_c$. From Figure \ref{fig:spectrum_Ml} we deduce that the
pattern is unstable for $S$ below some threshold where the dominant
eigenvalue of $\mathscr{M}_0$ equals zero. In fact, this threshold is
easily determined to correspond to $S_c=S_\text{crit}$, where
$\mu^{\prime}(S_\text{crit})=0$, since by differentiating the core
problem \eqref{eq:core} with respect to $S$ and comparing the
resulting system with \eqref{lin:psi_inner} when $l=0$, we conclude
that $B(0,S_c)=\mu^{\prime}(S_c)$. The dotted curve in Figure
\ref{fig:B_continuous} shows that the zero level curve
$B(\lambda,S_c)=0$ is such that $\lambda>0$ for
$S_c<S_\text{crit}$. As such, we conclude from \eqref{eq:Sc_equation}
that symmetric $N$-spot quasi-equilibria are unstable to competition
instabilities when
$\kappa > \kappa_{c1} \equiv \mu(S_\text{crit})/S_\text{crit}$.

For special spot configurations $\lbrace{x_1,\ldots,x_N\rbrace}$ where
$\pmb{e}$ is an eigenvector of $\mathcal{G}_0$ we can easily calculate
a higher order correction to this instability threshold. Since
$\mathcal{G}_0$ is symmetric, there are $N-1$ mutually
orthogonal eigenvectors $\pmb{q}_2,\ldots,\pmb{q}_N$ such that
$\mathcal{G}_0\pmb{q}_k=g_k\pmb{q}_k$ with
$\pmb{q}_k^T\pmb{e}=0$. Setting $\pmb{c}=\pmb{q}_k$ in
\eqref{eq:B_matching_D0}, and using
$B(0,S)\sim\varepsilon\mu^{\prime\prime}(S_\text{crit})\delta$ for $S=S_\text{crit}+
\varepsilon \delta$, we can determine the perturbed stability threshold
where $\lambda=0$ associated with each eigenvector $\pmb{q}_k$. By
taking the minimum of such values, and by recalling the
refined approximation \eqref{eq:Sc_eps_equation}, we obtain
that $N$-spot symmetric quasi-equilibria are all unstable on the range
\begin{equation}\label{lin:cyclic_symmetric}
  S_{c\varepsilon} < S_\text{crit} + \frac{4\pi\varepsilon}
  {\mu^{\prime\prime}(S_\text{crit})}\min_{k=2,\ldots,N}g_k \,.
\end{equation}

\begin{figure}[t!]
  \centering
	\begin{subfigure}[b]{0.33\textwidth}
		\centering
		\includegraphics[width=\textwidth,height=4.2cm]{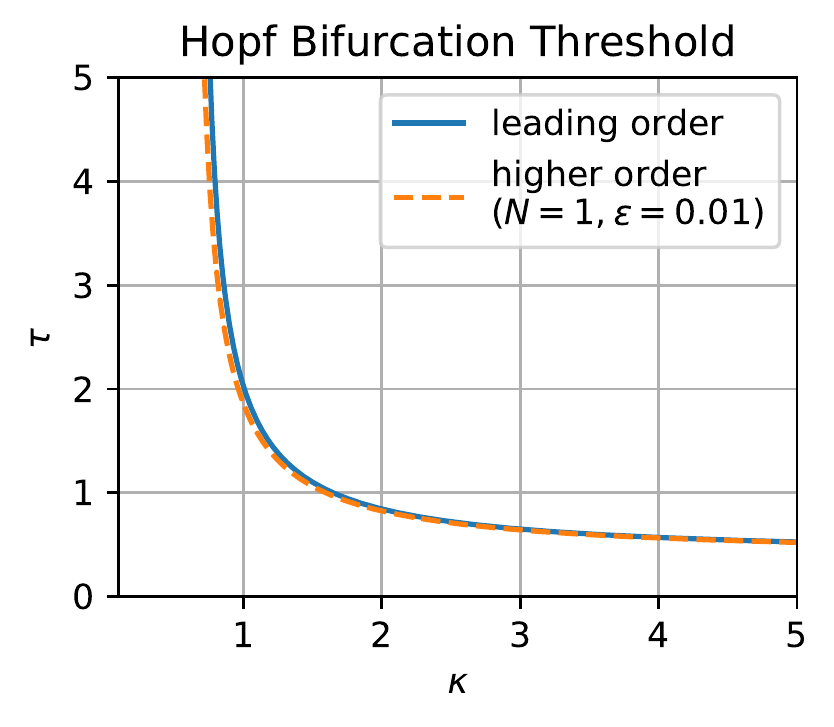}
		 \caption{}
                \label{fig:hopf_tau}
	\end{subfigure}%
	\begin{subfigure}[b]{0.33\textwidth}
		\centering
		\includegraphics[width=\textwidth,height=4.2cm]{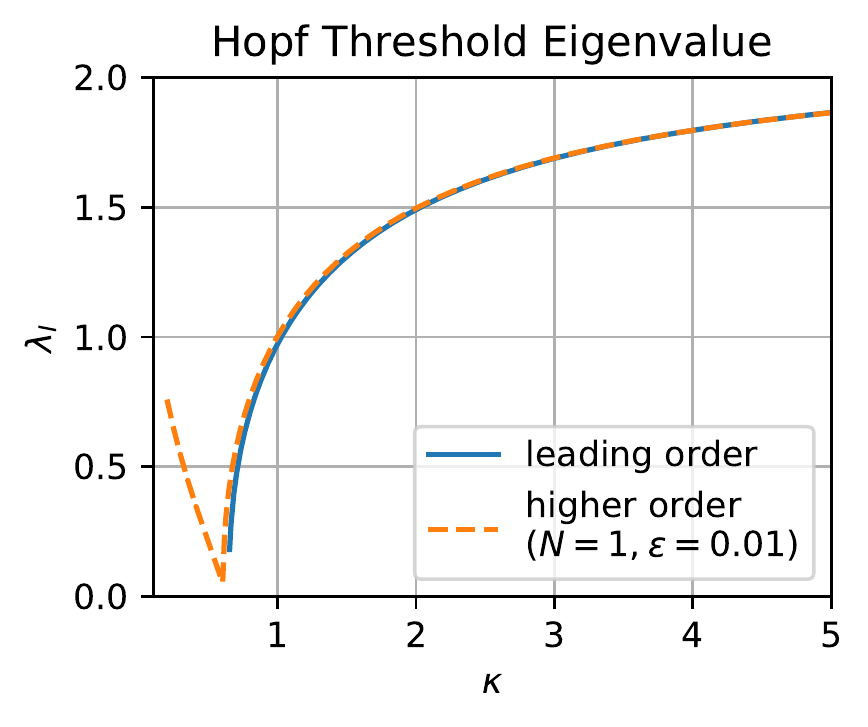}
		 \caption{}
                \label{fig:hopf_lambda}
	\end{subfigure}%
	\begin{subfigure}[b]{0.33\textwidth}
		\centering
		\includegraphics[width=\textwidth,height=4.2cm]{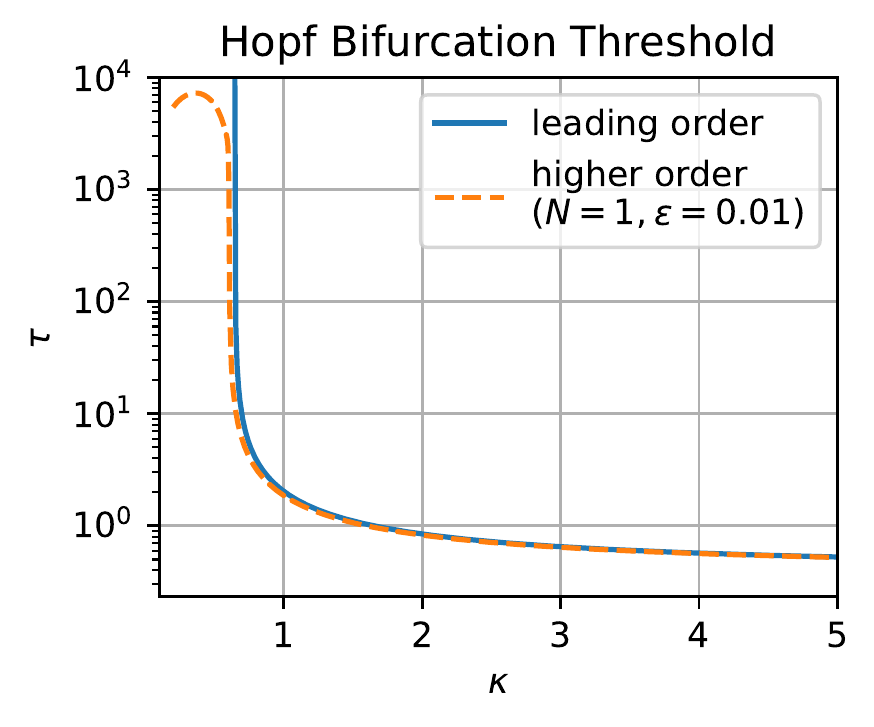}
		 \caption{}
                \label{fig:hopf_tau_large}
	\end{subfigure}
	\caption{Leading order (a) Hopf bifurcation threshold
          $\tau_h(\kappa)$ and (b) critical eigenvalue
          $\lambda = i\lambda_h$ for a symmetric $N$-spot pattern
          as calculated by solving \eqref{lin:hopf}
          numerically. The leading order theory
          assumes $\varepsilon|1+\tau\lambda|/D_0\ll1$ and is
          independent of the spot locations. We calculate the higher
          order Hopf bifurcation threshold for an $N=1$ spot pattern
          centered at the origin of the unit ball with
          $\varepsilon=0.01$ by solving \eqref{eq:B_matching} directly
          (note $\kappa = 3 D_0$). In (c) we see that although the
          leading order Hopf bifurcation threshold diverges as
          $\kappa\rightarrow\kappa_{c1}$, going to higher order
          demonstrates that a large but finite threshold
          persists.}\label{fig:hopf_threshold}
\end{figure}

Next we consider the case $\pmb{c} = \pmb{e}$ for which we find from
\eqref{eq:B_matching_D0} that, to leading order, $\lambda$ satisfies
\begin{equation}\label{eq:synchronous_leading_order}
B(\lambda,S_c) - \frac{\kappa}{1+\tau\lambda} = 0\,.
\end{equation}
First, we note that $\lambda=0$ is not a solution of
\eqref{eq:synchronous_leading_order} since, by using
$B(0,S)=\mu^{\prime}(S)$, this would require that
$\mu^{\prime}(S_c) = \kappa$, which the short argument following
\eqref{eq:symm_sol_D_large} demonstrates is impossible. Therefore, the
$\pmb{c}=\pmb{e}$ mode does not admit a zero-eigenvalue crossing and
any instability that arises must occur through a Hopf bifurcation. We
will seek a leading order threshold $\tau = \tau_h(\kappa)$ beyond
which a Hopf bifurcation is triggered. To motivate the existence of
such a threshold we consider first the $\kappa\rightarrow\infty$ limit
for which the asymptotics \eqref{eq:Sc_asymptotics} implies that
$S_c={1/(b\kappa^2)} \ll 1$ so that from the small $S$ expansion
\eqref{eq:core_sol_small_S} of the core solution we calculate from
\eqref{eq:ml_operator} that
$\mathscr{M}_0 \Phi \sim \Delta_\rho \Phi - \Phi + 2 w_c\Phi +
{\mathcal O}(\kappa^{-1})$. Then, by substituting this expression,
together with the small $S$ asymptotics \eqref{eq:core_sol_small_S}
where $S_c\sim {1/b\kappa^2}\ll 1$, into \eqref{eq:B_equation} we can
determine $B(\lambda,S_c)$ when $\kappa\gg 1$.  Then, by using the
resulting expression for $B$ in \eqref{eq:synchronous_leading_order},
we obtain the following well-known nonlocal eigenvalue problem (NLEP)
corresponding to the shadow limit
$\kappa={4\pi ND_0/|\Omega|} \to \infty$:
\begin{equation}
  1+\tau\lambda - \frac{2 \int_0^\infty w_c (\Delta_\rho - 1 + 2 w_c -
    \lambda)^{-1}w_c^2\rho^2 \, d\rho}{\int_0^\infty w_c^2\rho^2 \, d\rho} = 0 \,.
\end{equation}
>From Table 1 in \cite{ward_2003_hopf}, this NLEP has a Hopf
bifurcation at $\tau=\tau_h^\infty \approx 0.373$ with corresponding
critical eigenvalue $\lambda=i\lambda_h^\infty$ with
$\lambda_h^{\infty} \approx 2.174$.  To determine $\tau_h(\kappa)$ for
$\kappa={\mathcal O}(1)$, we set $\lambda=i\lambda_h$ in
\eqref{eq:synchronous_leading_order} and separate the resulting
expression into real and imaginary parts to obtain
\begin{equation}\label{lin:hopf}
  \tau_h = -\frac{\mbox{Im}\left(B(i\lambda_h,S_c)\right)}{\lambda_h
    \mbox{Re} \left(B(i\lambda_h,S_c)\right)}\,,\qquad\qquad
  \frac{|B(i\lambda_h,S_c)|^2}{\mbox{Re}\left(B(i\lambda_h,S_c)\right)} -
  \kappa = 0\,,
\end{equation}
where $S_c$ depends on $\kappa$ from \eqref{eq:Sc_equation}.  Starting
with $\kappa=50$ we solve the second equation for $\lambda_h$ using
Newton's method with $\lambda_h=\lambda_h^\infty$ as an initial
guess. We then use the first equation to calculate
$\tau_h$. Decreasing $\kappa$ and using the previous solution as an
initial guess we obtain the curves $\tau_h(\kappa)$ and
$\lambda_h(\kappa)$ as shown in Figure \ref{fig:hopf_threshold}.

We conclude this section by noting that as seen in Figures
\ref{fig:hopf_tau} and \ref{fig:hopf_tau_large} the leading order Hopf
bifurcation threshold diverges as $\kappa\rightarrow \kappa_{c1}^+$,
where $\kappa_{c1}={\mu(S_\text{crit})/S_\text{crit}}$. This is a
direct consequence of the assumption that
$\varepsilon|1+\tau\lambda|/D_0\ll1$ which fails to hold as $\tau$
gets increasingly large. Indeed, by using the series
  expansion in (3.12)--(3.14) of \cite{straube} for the reduced wave
  Green's function in the sphere, we can solve \eqref{eq:B_matching}
  directly using Newton's method for an $N=1$ spot configuration
  centered at the origin of the unit ball. Fixing $\varepsilon=0.001$,
  this yields the higher order asymptotic approximation for the Hopf
bifurcation threshold indicated by the dashed lines in Figure
\ref{fig:hopf_threshold}. This shows that to higher order the
bifurcation threshold is large but finite in the region
$\kappa \leq \kappa_{c1}$. Moreover, it hints at an $\varepsilon$
dependent rescaling of $\tau$ in the region $\kappa\leq \kappa_{c1}$
for which a counterpart to \eqref{eq:B_matching_D0_symmetric} may be
derived. While we do not undertake this rescaling in this paper we
remark that for 2-D spot patterns this rescaling led to the discovery
in \cite{tzou_2018_anomalous} of an \textit{anomalous} scaling law for
the Hopf threshold.

\subsubsection{Stability of Asymmetric Patterns in the $D=D_0/\varepsilon$ Regime}
When the $N$-spot pattern consists of $n$ \textit{large} spots of
strength $S_1=\ldots=S_n=S_r$ and $N-n$ \textit{small} spots of strength
$S_{n+1}=\ldots=S_N=S_l$, the leading order linear stability is
characterized by the blocked matrix system
\begin{equation}\label{eq:asymmetric_stability}
\begin{pmatrix}
B(\lambda,S_r)\mathcal{I}_{n} & 0 \\ 0 & B(\lambda,S_l)\mathcal{I}_{N-n}
\end{pmatrix}
\pmb{c} = \frac{\kappa}{1+\tau\lambda} \mathcal{E}_N \pmb{c}\,,
\end{equation}
where $\mathcal{I}_{m}$ denotes the $m\times m$ identity matrix.  In
particular, an asymmetric quasi-equilibrium solution is linearly
unstable if this system admits any nontrivial modes, $\pmb{c}$, for
which $\lambda$ has a positive real part. We will show that asymmetric
patterns are always unstable by explicitly constructing unstable
modes.

First, we assume that $1\leq n < N-1$ and we choose $\pmb{c}$ to be a
mode satisfying
\begin{equation}\label{eq:mode_case_1_unstable}
c_1 = \cdots = c_n = 0\,,\qquad c_{n+1} + \cdots + c_{N} = 0\,.
\end{equation}
Note that this mode describes \textit{competition} among the $N-n$
small spots of strength $S_l$. For such a mode,
\eqref{eq:asymmetric_stability} reduces to the single equation
$B(\lambda,S_l) = 0$, which implies that $\lambda$ must be an
eigenvalue of $\mathscr{M}_0$ at $S=S_l$. However, since
$S_l < S_\text{crit}$, we deduce from Figure \ref{fig:spectrum_Ml}
that there exists a real and positive $\lambda$ for $\mathscr{M}_0$ at
$S=S_l$. As such, any mode $\pmb{c}$ satisfying
\eqref{eq:mode_case_1_unstable} is linearly unstable.

We must consider the $n=N-1$ case separately since
\eqref{eq:mode_case_1_unstable} fails to yield nontrivial
modes. Instead of considering competition between the small spots, we
instead consider competition between large and small spots
collectively. We assume that $n \geq N-n$, for which $n=N-1$ is a
special case, and we try to exhibit an unstable mode $\pmb{c}$ of the form
\begin{equation}\label{eq:mode_case_2_unstable}
c_1=\ldots=c_n= c_r\,,\qquad c_{n+1}=\ldots=c_N = c_l\,.
\end{equation}
Then, \eqref{eq:asymmetric_stability} reduces to the system of two
equations
\begin{equation*}
  \left(B(\lambda,S_r) -\tfrac{\kappa}{1+\tau\lambda} \tfrac{n}{N} \right)c_r
  - \tfrac{\kappa}{1+\tau\lambda}\tfrac{(N-n)}{N} c_l = 0\,, \qquad
  -\tfrac{\kappa}{1+\tau\lambda} \tfrac{n}{N} c_r +
  \left(B(\lambda,S_l) -\tfrac{\kappa}{1+\tau\lambda} \tfrac{(N-n)}{N} \right)
  c_l = 0\,,
\end{equation*}
which admits a nontrivial solution if and only if the determinant of
this $2\times 2$ system vanishes. Therefore, to show that this mode is
unstable it suffices to prove that the zero-determinant condition, written
as
\begin{equation}\label{lin:fval}
  F(\lambda) \equiv B(\lambda,S_l)B(\lambda,S_r) -
  \frac{\kappa}{1+\tau\lambda}\biggl(\frac{n}{N}B(\lambda,S_l) +
  \frac{(N-n)}{N}B(\lambda,S_r)\biggr) = 0 \,,
\end{equation}
has a solution $\lambda>0$. To establish this, we first 
differentiate $\mu(S_r)=\mu(S_l)$ with respect to $S_r$ to obtain the
identity $\mu^{\prime}(S_l)S_l^{\prime}(S_r) = \mu^{\prime}(S_r)$. Combining
this result with $B(0,S)=\mu^{\prime}(S)$ we calculate that
\begin{equation}\label{lin:fzero}
  F(0) = \mu^{\prime}(S_l)\biggl[ \mu^{\prime}(S_r) -
  \kappa\frac{(N-n)}{N}\biggl(\frac{n}{(N-n)} + \frac{dS_l}{dS_r} \biggr)
  \biggr]   \,.
\end{equation}
Using $\mu^{\prime}(S_l)>0$ and $\mu^{\prime}(S_r)<0$ together with
$S_l^{\prime}(S_r) > -1$ (see Figure \ref{fig:Sl_prime}) and the
assumption $n/(N-n)\geq 1$, we immediately deduce that $F(0) < 0$. Next,
we let $\lambda_0>0$ be the dominant eigenvalue of $\mathscr{M}_0$
when $S=S_l$ (see Figure \ref{fig:spectrum_Ml}) so that
$B(\lambda_0,S_l) =0$. Then, from \eqref{lin:fval} we obtain
\begin{equation}
  F(\lambda_0) = -\frac{\kappa}{1+\tau\lambda_0}\frac{(N-n)}{N}
  B(\lambda_0,S_r)\,.
\end{equation}
However, since $\mathscr{M}_0$ at $S=S_r>S_\text{crit}$ has no
positive eigenvalues (see Figure \ref{fig:spectrum_Ml}), we deduce
that $B(\lambda,S_r)$ is of one sign for $\lambda\geq 0$ and,
furthermore, it must be negative since $B(0,S_r)=\mu^{\prime}(S_r)<0$
(see Figure \ref{fig:B_continuous} for a plot of $B$ showing both its
continuity and negativity for all $\lambda>0$ when
$S>S_\text{crit}$). Therefore, we have $F(\lambda_0) > 0$ and so,
combined with \eqref{lin:fzero}, by the intermediate value theorem it
follows that $F(\lambda)=0$ has a positive solution. We summarize
our leading order linear stability results in the following
proposition:

\begin{proposition}(Linear Stability): \label{prop:stability}
  Let $\varepsilon\ll 1$ and assume that
  $t\ll {\mathcal O}(\varepsilon^{-3})$. When $D={\mathcal O}(1)$, the
  $N$-spot symmetric pattern from Proposition
  \ref{prop:quasiequilibrium} is linearly stable. If
  $D=\varepsilon^{-1} D_0$ then the symmetric $N$-spot pattern from
  Proposition \ref{prop:quasiequilibrium} is linearly stable with
  respect to zero-eigenvalue crossing instabilities if
  $\kappa < \kappa_{c1}\equiv
  {\mu(S_\text{crit})/S_\text{crit}}\approx 0.64619$ and is unstable
  otherwise. Moreover, it is stable with respect to Hopf instabilities
  on the range $\kappa>\kappa_{c1}$ if $\tau < \tau_h(\kappa)$ where
  $\tau_h(\kappa)$ is plotted in Figure \ref{fig:hopf_tau}.  Finally,
  every asymmetric $N$-spot pattern in the $D=\varepsilon^{-1} D_0$
  regime is always linearly unstable.
\end{proposition}

%On the $D=\varepsilon^{-1} D_0$ regime, our numerical computations in
%Figure \eqref{fig:hopf_tau_large} for $N=1$ has suggested that
%$\tau_h(\kappa)\gg 1$ when $\kappa<\kappa_{c1}$.

\section{Slow Spot Dynamics}\label{sec:slow_dynamics}

A wide variety of singularly perturbed RD systems are known to exhibit
slow dynamics of multi-spot solutions in 2-D domains
(cf.~\cite{schnak2d}, \cite{chen}, \cite{trinh_2016},
\cite{ward_survey}). In this section we derive a system of ODE's which
characterize the motion of the spot locations $x_1,\ldots,x_N$ for the
3-D GM model on a slow time scale. Since the only $N$-spot patterns
that may be stable on an ${\mathcal O}(1)$ time scale are (to leading
order) symmetric we find that the ODE system reduces to a gradient
flow. We remark that both the derivation and final ODE system are
closely related to those in \cite{tzou_2017_schnakenberg} for the 3-D
Schnakenberg model.

The derivation of slow spot dynamics hinges on establishing a
solvability condition for higher order terms in the asymptotic
expansion in the inner region near each spot. As a result, we begin by
collecting higher order expansions of the limiting behaviour as
$|x-x_i|\rightarrow 0$ of the Green's functions $G(x,x_j)$ and
$G_0(x,x_j)$ that satisfy \eqref{eq:greens_function_pde} and
\eqref{eq:G0_equation}, respectively. In particular, we calculate that
\begin{subequations}
\begin{equation}\label{eq:greens_limit_high_order_1}
  G(x_i + \varepsilon y, x_j) \sim \begin{cases} G(x_i,x_j) +
    \varepsilon y \cdot \nabla_1 G(x_i,x_j)\,, & i\neq j\,, \\
    \frac{1}{4\pi\varepsilon \rho} + R(x_i) + \varepsilon y \cdot
    \nabla_1 R(x_i;x_i)\,, & i=j \,, \end{cases}\qquad \mbox{as}
  \quad |x - x_i|\rightarrow 0  \,,
\end{equation}
where $\rho=|y|$ and $\nabla_1R(x_i;x_i)\equiv \nabla_x R(x;x_1)\vert_{x=x_1}$.
Likewise, for the Neumann Green's function, we have
\begin{equation}\label{eq:greens_limit_high_order_2}
  G_0(x_i + \varepsilon y, x_j) \sim \frac{D_0}{\varepsilon |\Omega|} +
  \begin{cases} G_0(x_i,x_j) + \varepsilon y \cdot \nabla_1 G_0(x_i,x_j)\,, &
    i\neq j\,, \\
    \frac{1}{4\pi\varepsilon \rho} + R_0(x_i) + \varepsilon y \cdot
    \nabla_1 R_0(x_i;x_i)\,, & i=j \,, \end{cases}\qquad \mbox{as} \quad
  |x - x_i|\rightarrow 0 \,,
\end{equation}
\end{subequations}
where $\nabla_1$ again denotes the gradient with respect to the first
argument. We next extend the asymptotic construction of
quasi-equilibrium patterns in \S\ref{sec:quasi} by allowing the spot
locations to vary on a slow time scale. In particular, a dominant
balance in the asymptotic expansion requires that $x_i = x_i(\sigma)$
where $\sigma = \varepsilon^{3} t$. For $x$ near $x_i$ we introduce the
two term inner expansion
\begin{equation}\label{eq:two_term_inn}
  v \sim D V_i \sim D(V_{i\varepsilon}(\rho) + \varepsilon^2 V_{i2}(y) + \cdots)\,,
  \qquad u\sim DU_i\sim D\bigl(U_{i\varepsilon}(\rho) + \varepsilon^2 U_{i2}(y) +
  \cdots \bigr)\,,
\end{equation}
where we note the leading order terms are
$V_{i\varepsilon}(\rho) \equiv V(\rho,S_{i\varepsilon})$ and
$U_{i\varepsilon}(\rho) \equiv U(\rho,S_{i\varepsilon})$. By using
the chain rule we calculate $\partial_t V_i = - \varepsilon^2 x_i^{\prime}
(\sigma)\cdot\nabla_y V_i$ and $\partial_t U_i = -
\varepsilon^2 x_i^{\prime}(\sigma)\cdot\nabla_y U_i$. In this way,
upon substituting \eqref{eq:two_term_inn} into \eqref{eq:pde_gm_3d}
we collect the ${\mathcal O}(\varepsilon^2)$ terms to obtain that
$V_{i2}$ and $U_{i2}$ satisfy
\begin{subequations}
\begin{equation}\label{eq:Wi2_system}
  \mathscr{L}_{i\varepsilon}\pmb{W}_{i2} \equiv \Delta_y \pmb{W}_{i2} +
  \mathcal{Q}_{i\varepsilon}\pmb{W}_{i2} = -\pmb{f}_{i\varepsilon}\,, \qquad
  y\in \R^2 \,,
\end{equation}
where
\begin{equation}\label{eq:Wi2_system_definitions}
  \pmb{W}_{i2} \equiv \begin{pmatrix} V_{i2} \\ U_{i2}\end{pmatrix}\,,\qquad
  \pmb{f}_{i\varepsilon} \equiv
  \begin{pmatrix}
    \rho^{-1} V_{i\varepsilon}^{\prime}(\rho) \, x_i^{\prime}(\sigma)\cdot y  \\
    -D^{-1}U_{i\varepsilon}
  \end{pmatrix}\,,
  \qquad \mathcal{Q}_{i\varepsilon} \equiv
\begin{pmatrix}
  -1 + 2U_{i\varepsilon}^{-1} V_{i\varepsilon} & - U_{i\varepsilon}^{-2}V_{i\varepsilon}^{2}\\
  2 V_{i\varepsilon} & 0\end{pmatrix} \,.
\end{equation}
\end{subequations}
It remains to determine the appropriate limiting behaviour as
$\rho\rightarrow\infty$. From the first row of
$\mathcal{Q}_{i\varepsilon}$, we conclude that $V_{i2}\rightarrow 0$
exponentially as $\rho\rightarrow\infty$. However, the limiting
behaviour of $U_{i2}$ must be established by matching with the outer
solution. To perform this matching, we first use the distributional limit
\begin{equation*}
  \varepsilon^{-2} v^2 \longrightarrow 4\pi \varepsilon D^2 \sum_{j=1}^{N}
  S_{j\varepsilon}\delta(x-x_j) + 2\varepsilon^3 D^2 \sum_{j=1}^{N}
  \left(\int_{\mathbb{R}^3} V_{j\varepsilon}V_{j2} \, dy\right) \,  \delta(x-x_j)\,,
\end{equation*}
where the localization at each $x_1,\ldots,x_N$ eliminates all cross
terms. We then update \eqref{eq:u_outer} to include the
${\mathcal O}(\varepsilon^3)$ correction term. This leads to the refined
approximation for the outer solution
\begin{equation}
  u \sim 4\pi \varepsilon D \sum_{j=1}^{N} S_{j\varepsilon} G(x;x_j) +
  2\varepsilon^3 D \sum_{j=1}^{N} \left(\int_{\mathbb{R}^3} V_{j\varepsilon}V_{j2} \,
    dy \right) \, G(x;x_j) \,.
\end{equation}
We observe that the leading order matching condition is immediately
satisfied in both the $D={\mathcal O}(1)$ and the $D={D_0/\varepsilon}$
regimes. To establish the higher order matching condition we
distinguish between the $D={\mathcal O}(1)$ and
$D=\varepsilon^{-1}D_0$ regimes and use the higher order expansions of
the Green's functions as given by \eqref{eq:greens_limit_high_order_1}
and \eqref{eq:greens_limit_high_order_2}. In this way, in the
$D={\mathcal O}(1)$ regime we obtain the far-field behaviour as
$|y|\to \infty$ given by
\begin{equation}\label{eq:ffui2_D}
  U_{i2} \sim \frac{1}{2\pi\rho}\int_{\mathbb{R}^3}V_{i\varepsilon}V_{i2} \, dy +
  y\cdot b_{i\varepsilon} \,, \quad
 \frac{b_{i\varepsilon}}{4\pi}\equiv S_{i\varepsilon}\nabla_1 R(x_i;x_i) + \sum_{j\neq i}
  S_{j\varepsilon}\nabla_1 G(x_i,x_j) \,.
\end{equation}
Similarly, in the $D={D_0/\varepsilon}$ regime we obtain the following
far-field matching condition as $|y|\to\infty$:
\begin{equation}\label{eq:ffui2_D0}
  U_{i2} \sim \frac{1}{2\pi\rho}\int_{\mathbb{R}^3}V_{i\varepsilon}V_{i2} \,dy +
  \frac{2 D_0}{|\Omega|}\sum_{j=1}^{N} \int_{\mathbb{R}^3} V_{j\varepsilon}V_{j2}\,dy
  + y\cdot b_{0i\varepsilon}\,, \quad
  \frac{b_{0i\varepsilon}}{4\pi} \equiv S_{i\varepsilon}\nabla_1 R_0(x_i;x_i) +
  \sum_{j\neq i} S_{j\varepsilon}\nabla_1 G_0(x_i,x_j)\,.
\end{equation}
In both cases, our calculations below will show that only
$b_{i\varepsilon}$ and $b_{0i\varepsilon}$ affect the slow spot dynamics.

To characterize slow spot dynamics we calculate $x_i^{\prime}(\sigma)$
by formulating an appropriate solvability condition. We observe for
each $k=1,2,3$ that the functions
$\partial_{y_k}\pmb{W}_{i\varepsilon}$ where
$\pmb{W}_{i\varepsilon} \equiv (V_{i\varepsilon},U_{i\varepsilon})^T$
satisfy the homogeneous problem
$\mathscr{L}_{i\varepsilon}\partial_{y_k}\pmb{W}_{i\varepsilon} =
0$. Therefore, the null-space of the adjoint operator
$\mathscr{L}_{i\varepsilon}^\star$ is at least
three-dimensional. Assuming it is exactly three dimensional we
consider the three linearly independent solutions
$\pmb{\Psi}_{ik} \equiv y_k\pmb{P}_i(\rho)/\rho$ to the homogeneous
adjoint problem, where each
$\pmb{P}_i(\rho) = (P_{i1}(\rho), P_{i2}(\rho)^T$ solves
\begin{equation}\label{eq:P_problem}
  \Delta_\rho \pmb{P}_i - \frac{2}{\rho^2}\pmb{P}_i +
  \mathcal{Q}_{i\varepsilon}^T\pmb{P}_i = 0\,,\quad\rho>0\,;\quad
  \pmb{P}_i^{\prime}(0) = \begin{pmatrix} 0 \\ 0 \end{pmatrix} \,; \quad \mbox{with} \quad
  \mathcal{Q}_{i\varepsilon}^T \longrightarrow
  \begin{pmatrix} -1 & 0 \\ 0 & 0\end{pmatrix} \quad \mbox{as}\,\,\,
  \rho\to\infty \,.
\end{equation}
Owing to this limiting far-field behavior of the matrix
$\mathcal{Q}_{i\varepsilon}^T$, we immediately deduce that
$P_{i2} = {\mathcal O}(\rho^{-2})$ and that $P_{i1}$ decays
exponentially to zero as $\rho\rightarrow\infty$. Enforcing, for
convenience, the point normalization condition $P_{i2} \sim \rho^{-2}$
as $\rho\rightarrow \infty$, we find that \eqref{eq:P_problem} admits
a unique solution. We use each $\pmb{\Psi}_{ik}$ to impose a
solvability condition by multiplying \eqref{eq:Wi2_system} by
$\pmb{\Psi}_{ik}^T$ and integrating over the ball, $B_{\rho_0}$,
centered at the origin and of radius $\rho_0$ with $\rho_0 \gg
1$. Then, by using the divergence theorem, we calculate
\begin{equation}\label{eq:full_solve}
  \lim_{\rho_0\rightarrow\infty}\int_{B_{\rho_0}}\biggl(\pmb{\Psi}_{ik}^T\,
  \mathscr{L}_i \pmb{W}_{i2} - \pmb{W}_{i2} \, \mathscr{L}_i^\star
  \pmb{\Psi}_{ik}\biggr)\,dy = 
  \lim_{\rho_0\rightarrow\infty}\int_{\partial B_{\rho_0}}
  \biggl( \pmb{\Psi}_{ik}^T\partial_\rho 
\pmb{W}_{i2} - \pmb{W}_{i2}^T\partial_\rho\pmb{\Psi}_{ik} \biggr)
\biggr|_{\rho=\rho_0} \rho_0^2 \, d\Theta \,,
\end{equation}
where $\Theta$ denotes the solid angle for the unit sphere.

To proceed, we use the following simple identities given in
terms of the Kronecker symbol $\delta_{kl}$:
\begin{equation}\label{eq:ident_solve}
\int_{B_{\rho_0}} y_k f(\rho) \, dy = 0\, ,\qquad 
\int_{B_{\rho_0}}y_ky_lf(\rho)\, dy = \delta_{kl}\frac{4\pi}{3}
\int_0^{\rho_0}\rho^4 f(\rho) \, d\rho\,, \qquad \mbox{for} \quad
 l,k=1,2,3 \,.
\end{equation}
Since $\mathscr{L}_i^\star \pmb{\Psi}_{ik} = 0$, we can use
\eqref{eq:Wi2_system} and \eqref{eq:ident_solve}
to calculate the left-hand side of \eqref{eq:full_solve} as
\begin{equation}\label{eq:solve_left}
\begin{split}
  \lim_{\rho_0\rightarrow\infty}\int_{B_{\rho_0}}\pmb{\Psi}_{ik}^T\mathscr{L}_i
  \pmb{W}_{i2} dy & = \lim_{\rho_0\rightarrow\infty}\biggl( -
 \sum_{l=1}^3 x_{il}^{\prime}(\sigma) \int_{B_{\rho_0}} y_k y_l \frac{P_{i1}(\rho)
   V_{i\varepsilon}^{\prime}(\rho)}{\rho^2} \, dy  +
  \frac{1}{D}\int_{B_{\rho_0}} y_k \frac{P_{i2}(\rho)U_{i\varepsilon}(\rho)}{\rho}
         \, dy  \biggr) \\
  & = -\frac{4\pi}{3}x_{ik}^{\prime}(\sigma) \int_0^\infty P_{i1}(\rho)
   V_{i\varepsilon}^{\prime}(\rho)\rho^2 \, d\rho \,.
\end{split}
\end{equation} 
Next, in calculating the right-hand side of \eqref{eq:full_solve} by
using the far-field behavior \eqref{eq:ffui2_D} and
\eqref{eq:ffui2_D0}, we observe that only $b_{i\varepsilon}$ and
$b_{0i\varepsilon}$ terms play a role in the limit. In particular, in
the $D={\mathcal O}(1)$ regime we calculate in terms of the components
of $b_{i\varepsilon l}$ of the vector $b_{i\varepsilon}$, as given in
\eqref{eq:ffui2_D}, that
\begin{equation}\label{eq:solve_right}
\begin{split}
  & \lim_{\rho_0\rightarrow\infty}\int_{\partial B_{\rho_0}}
    \pmb{\Psi}_{ik}^T\partial_\rho\pmb{W}_{i2}\bigr|_{\rho=\rho_0}\rho_0^2\,d\Theta
    = \lim_{\rho_0\rightarrow\infty} \sum_{l=1}^{3} b_{i\varepsilon l}
    \int_{\partial B_{\rho_0}} \frac{y_ky_l}{\rho_0^2} \, d\Theta =
    \frac{4 \pi}{3} b_{i\varepsilon k}\,, \\
  & \lim_{\rho_0\rightarrow\infty}\int_{\partial B_{\rho_0}} \pmb{W}_{i2}^T\partial_\rho
    \pmb{\Psi}_{ik}\bigr|_{\rho=\rho_0} \rho_0^2\, d\Theta =
    -2\lim_{\rho_0\rightarrow\infty} \sum_{l=1}^3 b_{i\varepsilon l}
    \int_{\partial B_{\rho_0}} \frac{ y_ky_l}{\rho_0^2} \, d\Theta =
    -\frac{8\pi}{3} b_{i\varepsilon k} \,.
\end{split}
\end{equation}  
>From \eqref{eq:full_solve}, \eqref{eq:solve_left}, and \eqref{eq:solve_right},
we conclude for the $D={\mathcal O}(1)$ regime that
\begin{equation}\label{eq:ode_gamma}
  x_{ik}^{\prime}(\sigma) = -\frac{3}{\gamma(S_{i\varepsilon})} \,
  b_{i\varepsilon k}\,,
  \qquad \mbox{where} \qquad \gamma(S_{i\varepsilon}) \equiv
  \int_0^\infty P_{i1}(\rho) V_{i}^{\prime}(\rho,S_{i\varepsilon})\rho^2\, d\rho\,,
\end{equation}
which holds for each component $k=1,2,3$ and each spot
$i=1,\ldots,N$. From symmetry considerations we see that the constant
contribution to the far-field behaviour, as given by the first term in
\eqref{eq:ffui2_D}, is eliminated when integrated over the
boundary. In an identical way, we can determine $x_{ik}^{\prime}$ for
the $D={D_0/\epsilon}$ regime.  In summary, in terms of the gradients
of the Green's functions and
$\gamma_{i\varepsilon}\equiv \gamma(S_{i\varepsilon})$, as defined in
\eqref{eq:ode_gamma}, we obtain the following vector-valued ODE
systems for the two distinguished ranges of $D$:
\begin{equation}\label{eq:ode_eps}
\begin{split}
  \frac{d x_i}{d\sigma} = -\frac{12\pi}{\gamma_{i\varepsilon}}
\begin{cases}
 \biggl( S_{i\varepsilon}\nabla_1 R(x_i;x_i) + \sum_{j\neq i} S_{j\varepsilon}
\nabla_1 G(x_i,x_j) \biggr)\,, & \quad \mbox{for} \,\,\, D={\mathcal O}(1)\,,\\
\biggl( S_{i\varepsilon}\nabla_1 R_0(x_i;x_i) + \sum_{j\neq i} S_{j\varepsilon}\nabla_1
G_0(x_i,x_j) \biggr)\,, & \quad \mbox{for} \,\,\, D={D_0/\varepsilon} \,.
\end{cases}
\end{split}
\end{equation}

\begin{figure}[t!]
	\centering
	\includegraphics[width=0.40\textwidth,height=4.2cm]{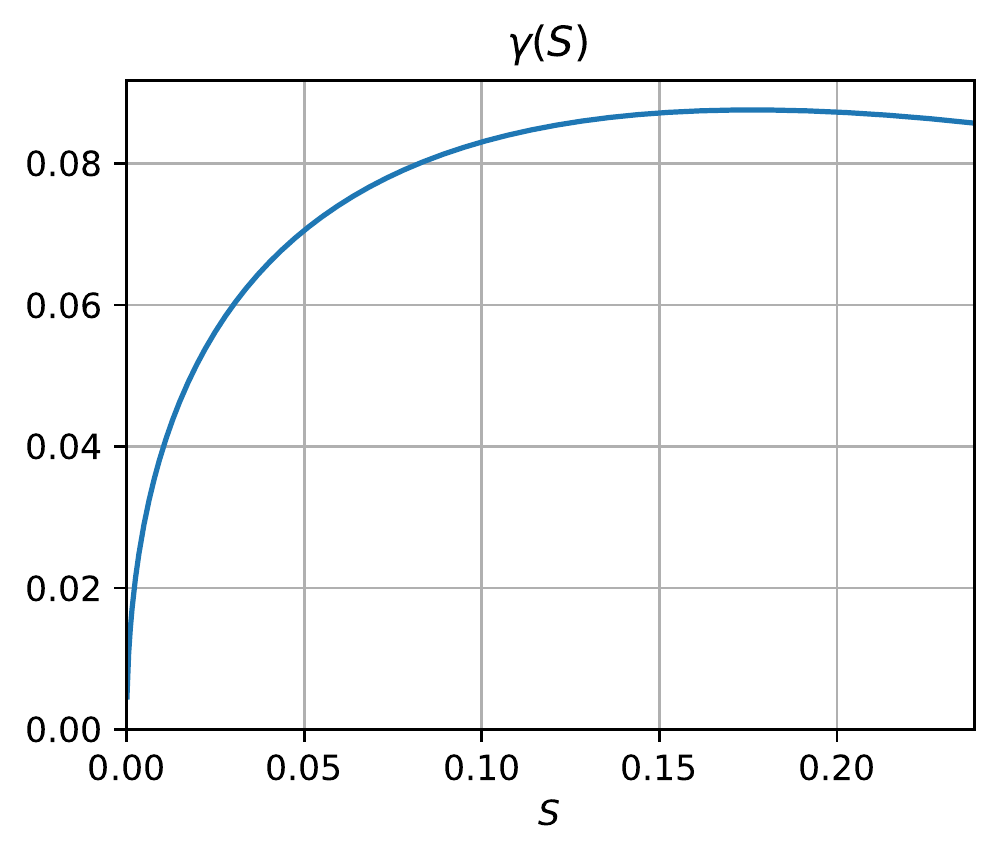}
	\caption{Plot of the numerically-computed multiplier $\gamma(S)$
          as defined in the slow gradient flow dynamics
          \eqref{eq:slow_dynamics_gradient_flow}.}\label{fig:gamma}
\end{figure}

Since only the symmetric $N$-spot configurations can be stable on an
${\mathcal O}(1)$ time scale (see Proposition \ref{prop:stability}),
it suffices to consider the ODE systems in \eqref{eq:ode_eps} when
$S_{i\varepsilon} = S_\star+{\mathcal O}(\varepsilon)$ in the
$D={\mathcal O}(1)$ regime and when
$S_{i\varepsilon} = S_c+{\mathcal O}(\varepsilon)$, where $S_c$ solves
\eqref{eq:Sc_equation}, in the $D=\varepsilon^{-1}D_0$ regime. In
particular, we find that to leading order, where the ${\mathcal O}(\varepsilon)$
corrections to the source strengths are neglected, the ODE systems in
\eqref{eq:ode_eps} can be reduced to the gradient flow dynamics
\begin{subequations}\label{eq:slow_dynamics_gradient_flow}
\begin{equation}
  \frac{d x_i}{d\sigma} = -\frac{6\pi S}{\gamma(S)}\nabla_{x_i}
  \mathscr{H}(x_1,\ldots,x_N) \,, \qquad \mbox{with} \quad
  \gamma(S) = \int_{0}^{\infty} P_1(\rho) V_1(\rho,S) \rho^2 \, d\rho \,,
\end{equation}
where $S=S_\star$ or $S = S_c$ depending on whether
$D={\mathcal O}(1)$ or $D=\varepsilon^{-1}D_0$, respectively. In
\eqref{eq:slow_dynamics_gradient_flow} the discrete energy
$\mathscr{H}$, which depends on the instantaneous spot locations, is
defined by
\begin{equation}
  \mathscr{H}(x_1,\ldots,x_N) \equiv \begin{cases} \sum_{i=1}^{N} R(x_i) +
   2 \sum_{i=1}^N \sum_{j>i} G(x_i,x_j)\,, & \quad \mbox{for} \quad
    D={\mathcal O}(1)\,, \\
    \sum_{i=1}^{N} R_0(x_i) + 2 \sum_{i=1}^N \sum_{j>i} G_0(x_i,x_j)\,,
    & \quad \mbox{for} \quad D=\varepsilon^{-1}D_0\,.\end{cases}
\end{equation}
\end{subequations}
In accounting for the factor of two between
\eqref{eq:slow_dynamics_gradient_flow} and \eqref{eq:ode_eps}, we used
the reciprocity relations for the Green's functions. In this leading
order ODE system, the integral $\gamma(S)$ is the same for each spot,
since $P_1(\rho)$ is computed numerically from the homogeneous adjoint
problem \eqref{eq:P_problem} using the core solution
$V_1(\rho,S)$ and $U_1(\rho,S)$ to calculate the matrix
$\mathcal{Q}_{i\varepsilon}^T$ in \eqref{eq:P_problem}. In Figure
\ref{fig:gamma} we plot the numerically-computed $\gamma(S)$, where we
note that $\gamma(S)>0$. Since $\gamma(S)>0$, local minima of $\mathscr{H}$
are linearly stable equilibria for
\eqref{eq:slow_dynamics_gradient_flow}.

We remark that this gradient flow system
\eqref{eq:slow_dynamics_gradient_flow} differs from that derived in
\cite{tzou_2017_schnakenberg} for the 3-D Schnakenberg model only
through the constant $\gamma(S)$. Since this parameter affects only
the time-scale of the slow dynamics we deduce that the equilibrium
configurations and stability properties for the ODE dynamics will be
identical to those of the Schnakenberg model. As such, we do not
analyze \eqref{eq:slow_dynamics_gradient_flow} further and instead
refer to \cite{tzou_2017_schnakenberg} for more detailed numerical
investigations. Finally we note that the methods employed here and in
\cite{tzou_2017_schnakenberg} should be applicable to other 3-D RD systems
yielding similar limiting ODE systems for slow spot dynamics.  The
similarity between slow dynamics for a variety of RD systems in 2-D has
been previously observed and a general asymptotic framework has been
pursued in \cite{trinh_2016} for the dynamics on the sphere.

\section{Numerical Examples}\label{sec:examples}

\begin{figure}[t!]
	\centering
	\begin{subfigure}{0.45\textwidth}
		\centering
		\includegraphics[scale=0.70]{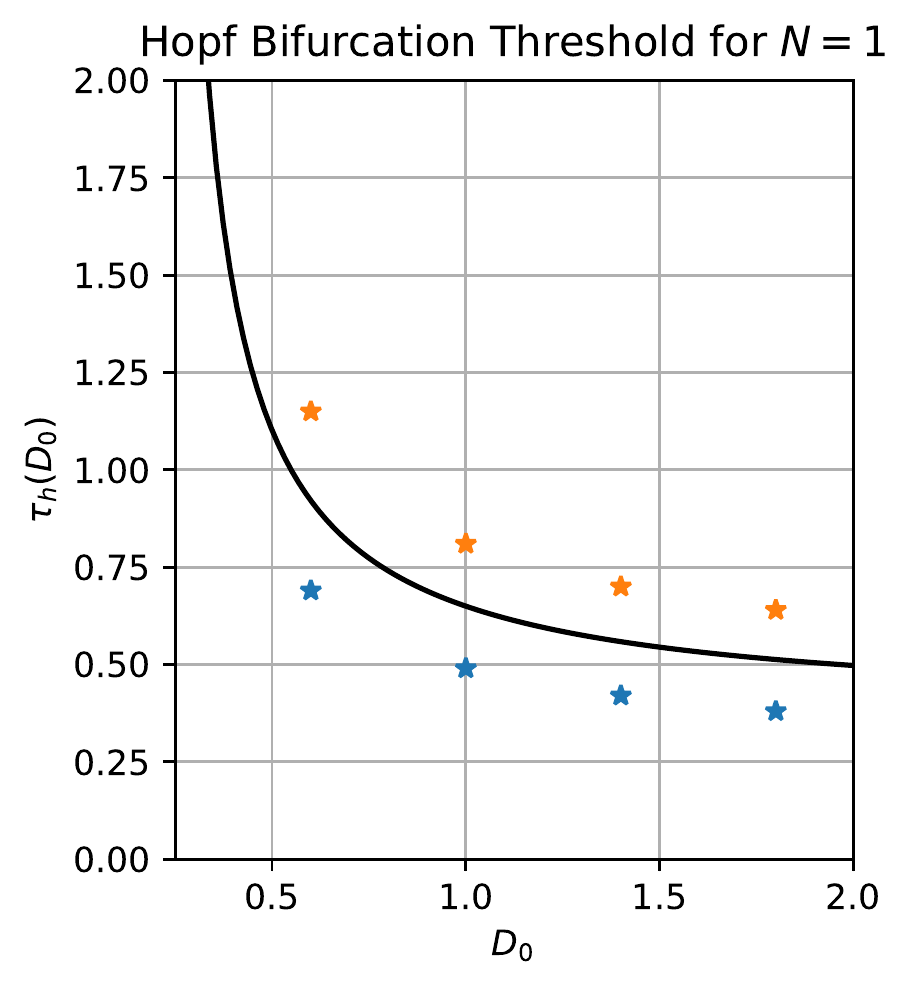}
		 \caption{}
                \label{fig:hopf_simulation_legend}
	\end{subfigure}%
	\begin{subfigure}{0.55\textwidth}
		\centering
		\includegraphics[scale=0.70]{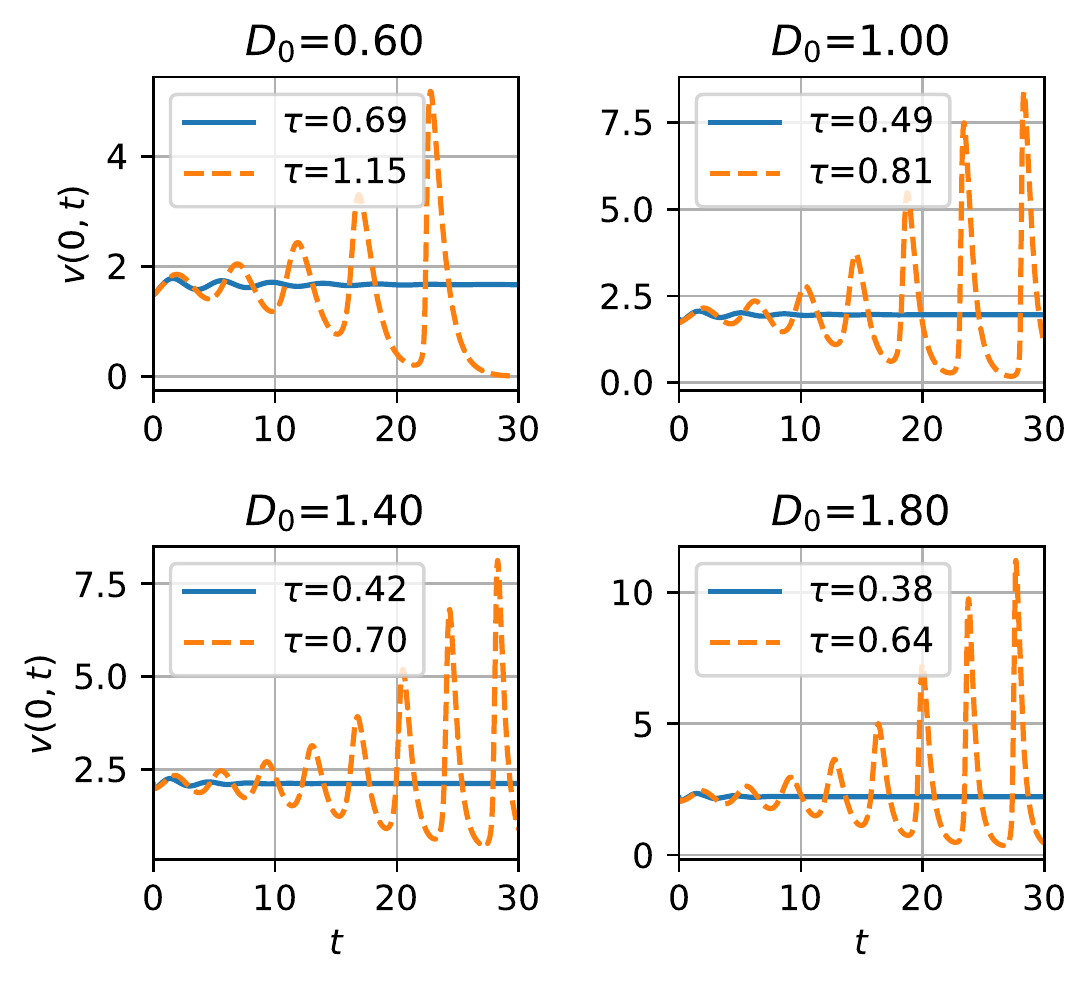}
		 \caption{}
                \label{fig:hopf_simulation_heights}
              \end{subfigure}%
              \vspace*{-0.5cm}
	\caption{(a) Leading order Hopf bifurcation threshold for a
          one-spot pattern. (b) Plots of the spot height $v(0,t)$
          from numerically solving \eqref{eq:pde_gm_3d} using
          FlexPDE6 \cite{flexpde} in the unit ball with
          $\varepsilon=0.05$ at the indicated $\tau$ and $D_0$
          values.}\label{fig:hopf_simulation}
\end{figure}

In this section we use FlexPDE6 \cite{flexpde} to numerically solve
\eqref{eq:pde_gm_3d} when $\Omega$ is the unit ball. In particular, we
illustrate the emergence of Hopf and competition instabilities, as
predicted in \S\ref{sec:stability} for symmetric spot patterns in the
$D={D_0/\varepsilon}$ regimes.

We begin by considering a single spot centered at the origin in the
unit ball, for the $D=\varepsilon^{-1}D_0$ regime. Since no
competition instabilities occur for a single spot solution, we focus
exclusively on the onset of Hopf instabilities as $\tau$ is
increased. In Figure \ref{fig:hopf_simulation_legend} we plot the Hopf
bifurcation threshold obtained from our linear stability theory, and
indicate several sample points below and above the threshold. Using
FlexPDE6 \cite{flexpde}, we performed full numerical simulations of
\eqref{eq:pde_gm_3d} in the unit ball with $\varepsilon=0.05$ and
parameters $D_0$ and $\tau$ corresponding to the labeled points in Figure
\ref{fig:hopf_simulation_legend}. The resulting activator height at
the origin, $v(0,t)$, computed from FlexPDE6 is shown in Figure
\ref{fig:hopf_simulation_heights} for these indicated parameter
values. We observe that there is good agreement with the onset of Hopf
bifurcations as predicted by our linear stability theory.

\begin{figure}[t!]
	\centering
	\begin{subfigure}{0.5\textwidth}
		\centering
		\includegraphics[scale=0.70]{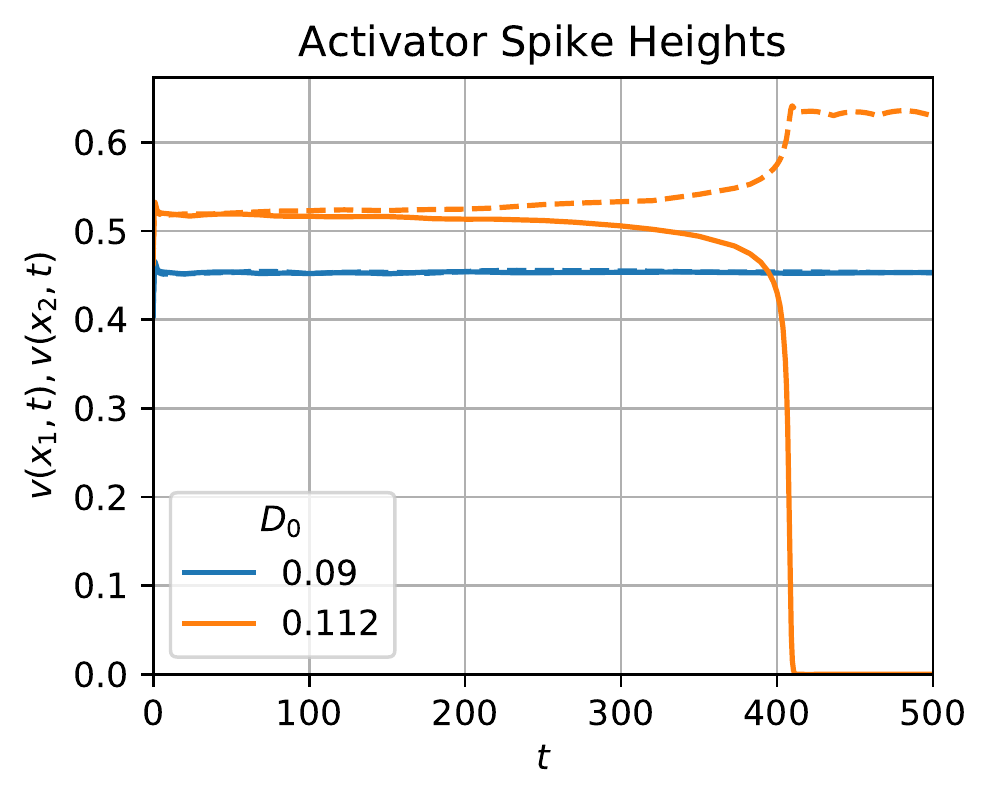}
		\caption{}
                \label{fig:competition_heights}
	\end{subfigure}%
	\begin{subfigure}{0.5\textwidth}
		\centering
		\includegraphics[scale=0.140]{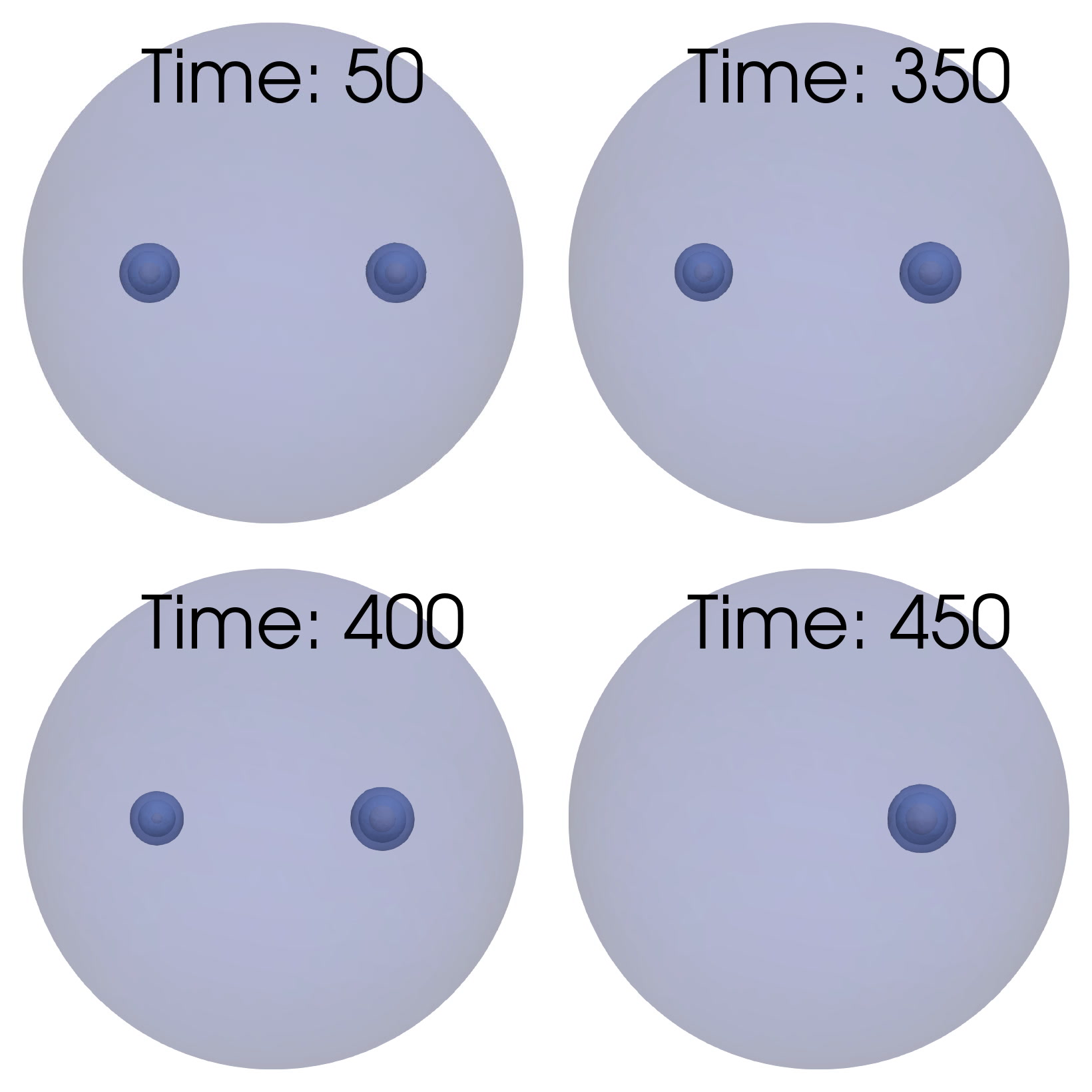}
		 \caption{}
                \label{fig:competition_snaps}
              \end{subfigure}%
                            \vspace*{-0.5cm}
	\caption{(a) Plots of the spot heights (solid and dashed
          lines) in a two-spot symmetric pattern at the indicated values
          of $D_0$. Results were obtained by using FlexPDE6
          \cite{flexpde} to solve \eqref{eq:pde_gm_3d} in the unit
          ball with $\varepsilon=0.05$ and $\tau = 0.2$. (b) 
          plot of three-dimensional contours of $v(x,t)$ for
          $D_0 = 0.112$, with contours chosen at $v=0.1,0.2,0.4$.}
          \label{fig:competition}
\end{figure}

Next, we illustrate the onset of a competition instability by
considering a symmetric two-spot configurations with spots centered at
$(\pm0.51565,0,0)$ in the unit ball and with $\tau=0.2$ (chosen small
enough to avoid Hopf bifurcations) and $\varepsilon=0.05$. The
critical value of $\kappa_{c1}\approx 0.64619$ then implies that the
leading order competition instability threshold for the unit ball with
$|\Omega|={4\pi/3}$ is $D_0 \approx 0.64619/(3N) = 0.108$. We
performed full numerical simulations of \eqref{eq:pde_gm_3d} using
FlexPDE6 \cite{flexpde} with values of $D_0 = 0.09$ and $D_0 =
0.112$. The results of our numerical simulations are shown in Figure
\ref{fig:competition}, where we observe that a competition instability
occurs for $D_0 = 0.112$, as predicted by the linear stability
theory. Moreover, in agreement with previous studies of competition
instabilities (cf.~\cite{tzou_2017_schnakenberg}, \cite{chen}), we
observe that a competition instability triggers a nonlinear event
leading to the annihilation of one spot.
 
\section{The Weak Interaction Limit $D={\mathcal O}(\varepsilon^2)$}\label{sec:weak}

\begin{figure}[t!]
	\centering
	\begin{subfigure}[b]{0.40\textwidth}
		\centering
		\includegraphics[width=\textwidth, height=4.2cm]{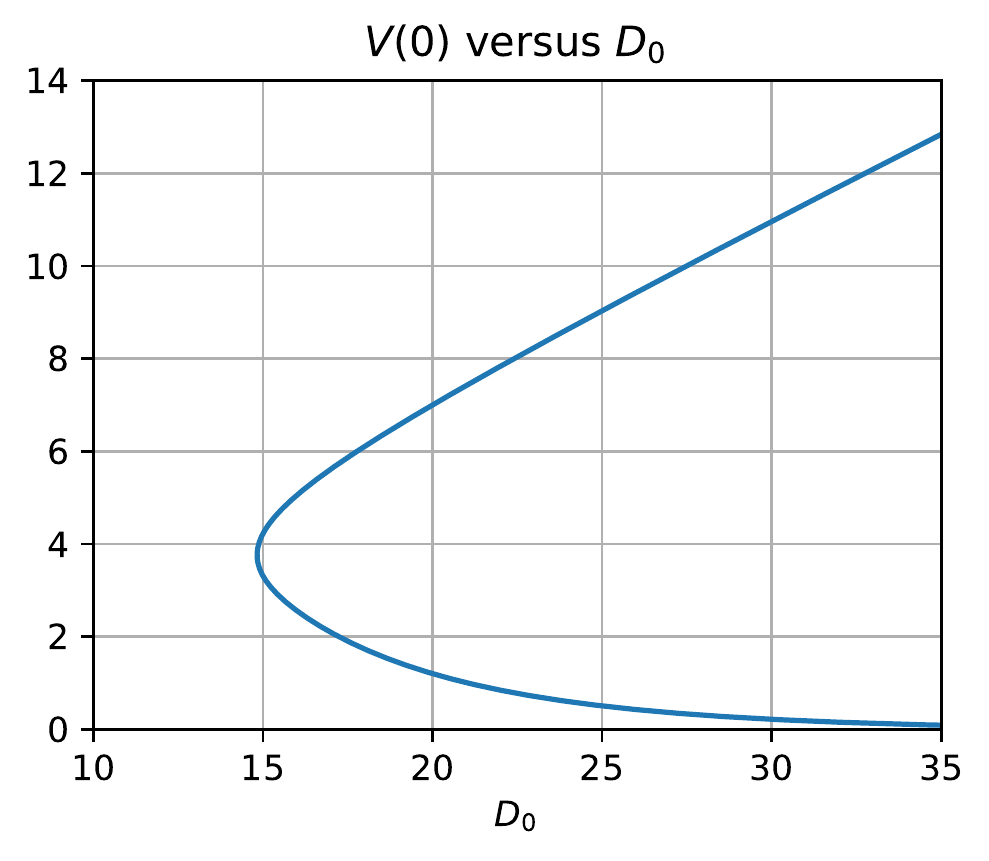}
		\caption{}
                \label{fig:eps2_regime}
	\end{subfigure}%
	\begin{subfigure}[b]{0.40\textwidth}
		\centering
		\includegraphics[width=\textwidth, height=4.2cm]{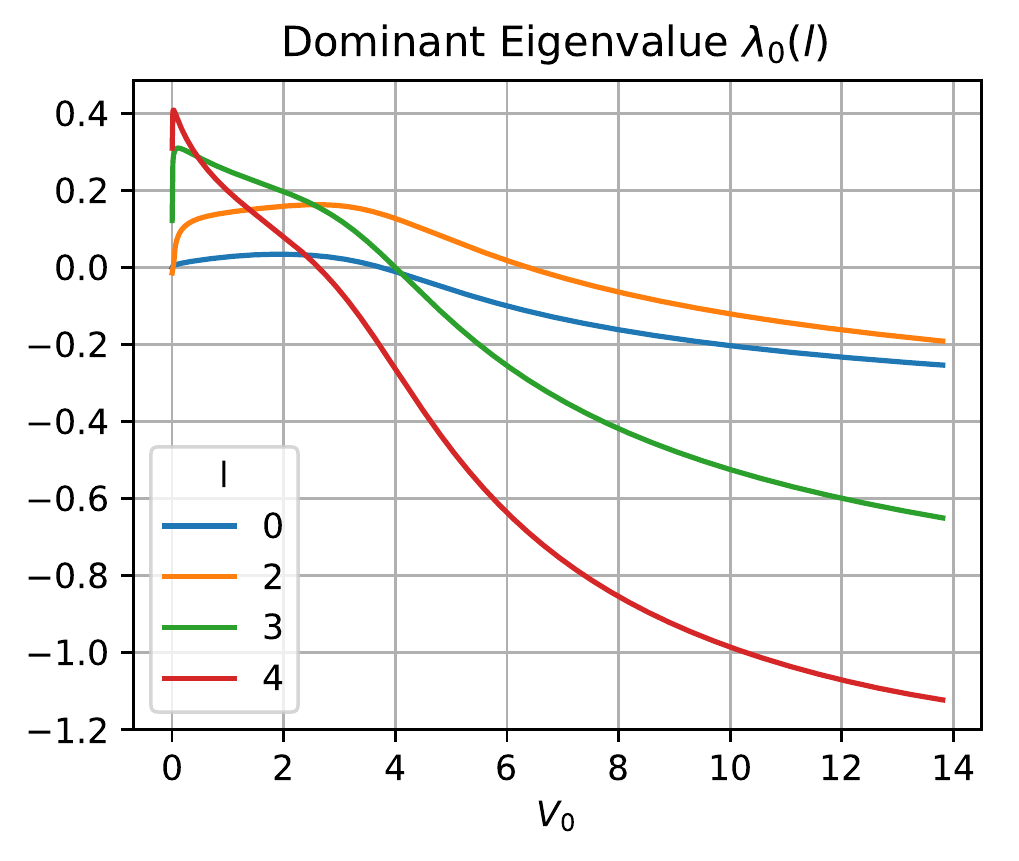}
		 \caption{}
                \label{fig:eps2_stability}
	\end{subfigure}%
	\caption{(a) Bifurcation diagram for solutions to the core
          problem \eqref{eq:core_eps2} in the $D=\varepsilon^2D_0$ regime. (b)
          Dominant eigenvalue of the linearization of the core problem for
          each mode $l=0,2,3,4$, as computed numerically from
         \eqref{eq:Ml_def}. }\label{fig:eps2}
\end{figure}

In \S\ref{sec:stability} we have shown in both the $D={\mathcal O}(1)$
and $D={\mathcal O}(\varepsilon^{-1})$ regimes that $N$-spot
quasi-equilibria are not susceptible to locally non-radially symmetric
instabilities. Here we consider the weak-interaction regime
$D=D_0\varepsilon^2$, where we numerically determine that locally
non-radially symmetric instabilities of a localized spot are possible.
First, we let $\xi\in\Omega$ satisfy
$\text{dist}(\xi,\partial\Omega)\gg {\mathcal O}(\varepsilon^2)$ and
we introduce the local coordinates $x = \xi + \varepsilon y$ and the
inner variables $v\sim \varepsilon^2 V(\rho)$ and
$u \sim \varepsilon^2U(\rho)$. With this scaling, and with
$D=D_0\varepsilon^2$, the steady-state problem for
\eqref{eq:pde_gm_3d} becomes
\begin{equation}\label{eq:core_eps2}
  \Delta_\rho V - V + U^{-1}V^2 = 0\,,\qquad D_0 \Delta_\rho U - U + V^2 = 0\,,
  \qquad \rho =|y| > 0 \,.
\end{equation}
For this core problem, we impose the boundary conditions
$V_\rho(0) = U_\rho(0) = 0$ and $(V,U)\rightarrow 0$ exponentially as
$\rho\rightarrow \infty$. Unlike the $D={\mathcal O}(1)$ and
$D={\mathcal O}(\varepsilon^{-1})$ regimes, $u$ and $v$ are both
exponentially small in the outer region. Therefore, for any
well-separated configuration $x_1,\ldots,x_N$, the inner problems near
each spot centre are essentially identical and independent. In Figure
\ref{fig:eps2_regime} we plot $V(0)$ versus $D_0$ obtained by
numerically solving \eqref{eq:core_eps2}. From this figure, we observe
that for all $D_0 \gtrapprox 14.825$, corresponding to a saddle-node
point, the core problem \eqref{eq:core_eps2} admits two distinct
radially-symmetric solutions.

Since both the activator $V$ and inhibitor $U$ decay exponentially there are
only exponentially weak interactions between individual spots. As a result,
it suffices to consider only the linear stability of the core problem
\eqref{eq:core_eps2}. Upon linearizing \eqref{eq:pde_gm_3d} about the
core solution we obtain the eigenvalue problem
\begin{equation}\label{new_eq:linear_stability_pde}
  \Delta_\rho\Phi - \frac{l(l+1)}{\rho^2}\Phi - \Phi + \frac{2V}{U}\Phi -
  \frac{V^2}{U^2}\Psi = \lambda \Phi\,,\qquad D_0 \Delta_\rho\Psi -
  \frac{l(l+1)}{\rho^2}\Psi - \Psi + 2V\Phi = 0\,,
\end{equation}
for each $l\geq 0$ and for which we impose that
$\Phi^{\prime}(0)=\Psi^{\prime}(0)=0$
and $(\Phi,\Psi)\rightarrow 0$ exponentially as
$\rho\rightarrow \infty$. We reduce
\eqref{new_eq:linear_stability_pde} to a single nonlocal equation by
noting that the Green's function $G_l(\rho,\rho_0)$ satisfying
\begin{equation}
  D_0\Delta_\rho G_l - \frac{l(l+1)}{\rho^2} G_l - G_l = -
  \frac{\delta(\rho-\rho_0)}{\rho^2}\,,
\end{equation}
is given explicitly by
\begin{equation}
  G_l(\rho,\rho_0) = \frac{1}{D_0\sqrt{\rho_0\rho}}
  \begin{cases} I_{l+1/2}(\rho/\sqrt{D_0}) K_{l+1/2}(\rho_0/\sqrt{D_0})\,,
    & \quad \rho<\rho_0\,,\\
    I_{l+1/2}(\rho_0/\sqrt{D_0}) K_{l+1/2}(\rho/\sqrt{D_0})\,, & \quad \rho>\rho_0
    \,,\end{cases}
\end{equation}
where $I_{n}(\cdot)$ and $K_{n}(\cdot)$ are the $n^\text{th}$ order
modified Bessel Functions of the first and second kind,
respectively. As a result, by proceeding as in \S\ref{sec:stability}
we reduce \eqref{new_eq:linear_stability_pde} to the nonlocal spectral
problem $\mathscr{M}_l\Phi = \lambda \Phi$ where
\begin{equation}\label{eq:Ml_def}
  \mathscr{M}_l\Phi \equiv \Delta_\rho \Phi -\frac{l(l+1)}{\rho^2}\Phi -
  \Phi + \frac{2V}{U}\Phi - \frac{2V^2}{U^2}\int_0^\infty
  G_l(\rho,\tilde{\rho})\,V(\tilde{\rho})\, \Phi(\tilde{\rho})\tilde{\rho}^2 \,
  d\tilde{\rho} \,.
\end{equation}
In Figure \ref{fig:eps2_stability} we plot the real part of the
largest numerically-computed eigenvalue of $\mathscr{M}_l$ as a
function of $V(0)$ for $l=0,2,3,4$. From this figure, we observe that
the entire lower solution branch in the $V(0)$ versus $D_0$
bifurcation diagram in Figure \ref{fig:eps2_regime} is
unstable. However, in contrast to the $D={\mathcal O}(1)$ and
$D={\mathcal O}(\varepsilon^{-1})$ regimes, we observe from the orange
curve in Figure \ref{fig:eps2_stability} for the $l=2$ mode that when
$D=\varepsilon^2 D_0$ there is a range of $D_0$ values for which a
peanut-splitting instability is the only unstable mode.

\begin{figure}[t!]
	\centering
	\begin{subfigure}{0.25\textwidth}
		\centering
		\includegraphics[scale=0.15]{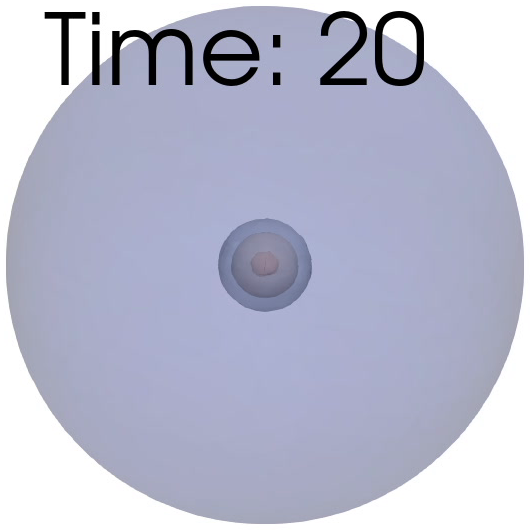}
		%		\caption{}\label{fig:splitting_snap_1}
	\end{subfigure}%
	\begin{subfigure}{0.25\textwidth}
		\centering
		\includegraphics[scale=0.15]{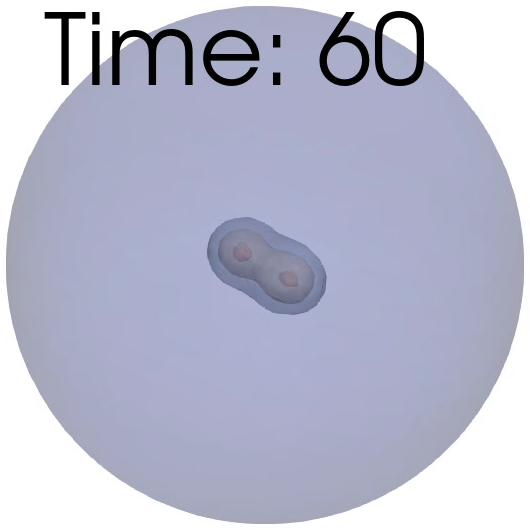}
		%		\caption{}\label{fig:splitting_snap_2}
	\end{subfigure}%
	\begin{subfigure}{0.25\textwidth}
		\centering
		\includegraphics[scale=0.15]{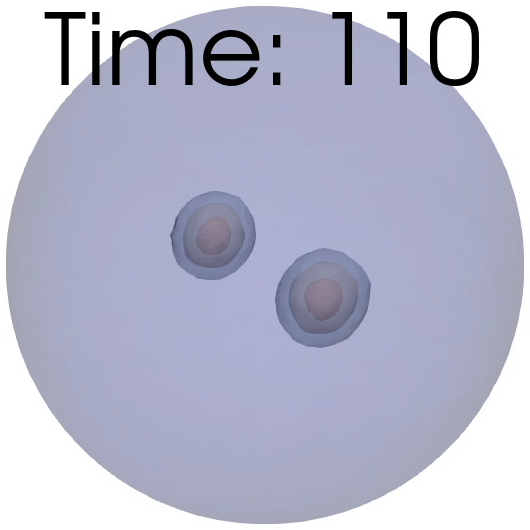}
		%		\caption{}\label{fig:splitting_snap_3}
	\end{subfigure}%
	\begin{subfigure}{0.25\textwidth}
		\centering
		\includegraphics[scale=0.15]{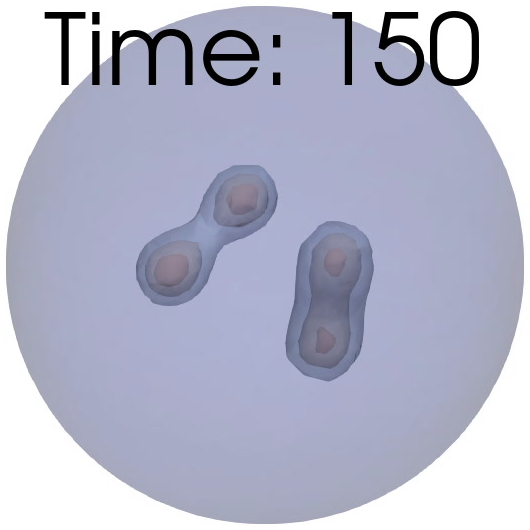}
		%		\caption{}\label{fig:splitting_snap_4}
	\end{subfigure}\\[0.80ex]%
	\begin{subfigure}{0.25\textwidth}
		\centering
		\includegraphics[scale=0.15]{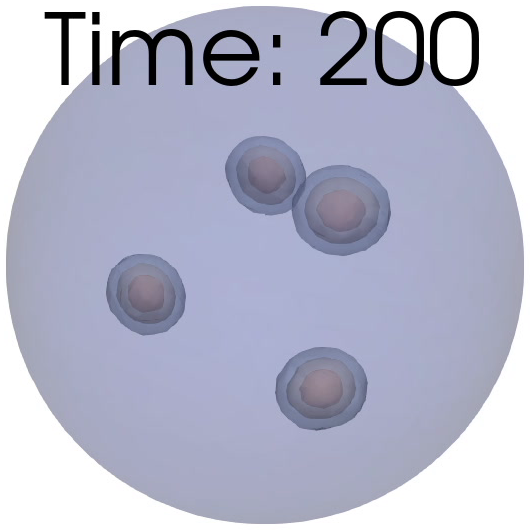}
		%		\caption{}\label{fig:splitting_snap_5}
	\end{subfigure}%
	\begin{subfigure}{0.25\textwidth}
		\centering
		\includegraphics[scale=0.15]{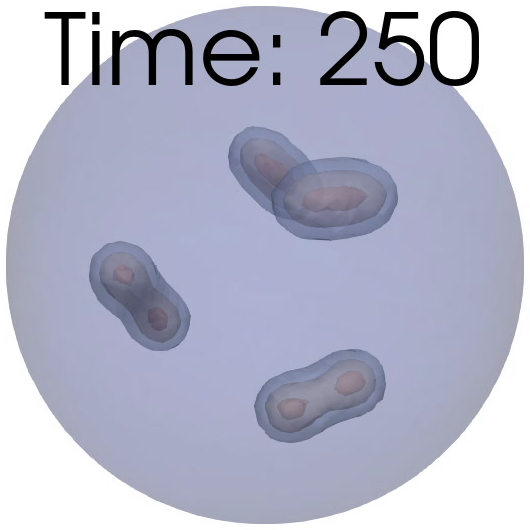}
		%		\caption{}\label{fig:splitting_snap_6}
	\end{subfigure}%
	\begin{subfigure}{0.25\textwidth}
		\centering
		\includegraphics[scale=0.15]{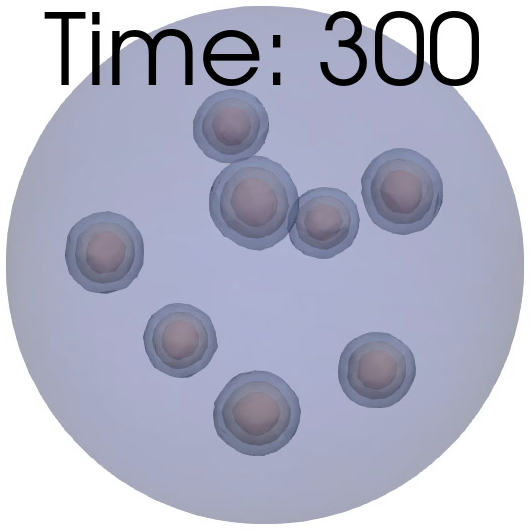}
		%		\caption{}\label{fig:splitting_snap_7}
	\end{subfigure}%
	\begin{subfigure}{0.25\textwidth}
		\centering
		\includegraphics[scale=0.15]{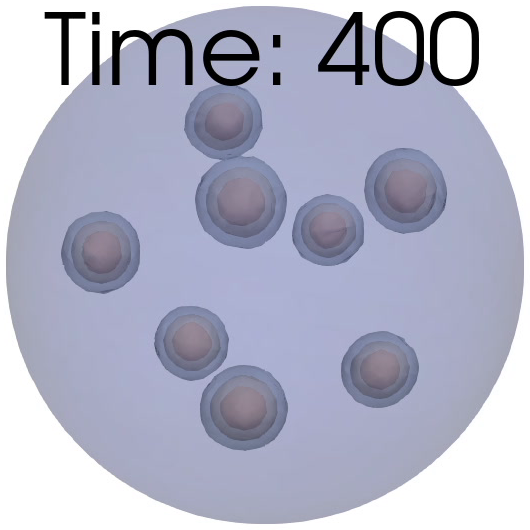}
		%		\caption{}\label{fig:splitting_snap_7}
	\end{subfigure}%
	\caption{Snapshots of FlexPDE6 \cite{flexpde} simulation of
          \eqref{eq:pde_gm_3d} in the unit ball with
          $\varepsilon=0.05$, $D=16\varepsilon^2$, and $\tau=1$ and
          with initial condition given by a single spot solution in
          the weak interaction limit calculated from
          \eqref{eq:core_eps2} with $V(0)=5$. The snapshots show
          contour plots of the activator $v(x,t)$ at different times
          where for each spot the outermost, middle, and innermost
          contours correspond to values of $0.006$, $0.009,$ and
          $0.012$ respectively. Note that the asymptotic theory
          predicts a maximum peak height of
          $v\sim \varepsilon^2 V(0) \approx
          0.0125$.}\label{fig:splitting}
\end{figure}

In previous studies of singularly perturbed RD systems supporting
peanut-splitting instabilities it has typically been observed that
such linear instabilities trigger nonlinear spot self-replication
events (cf.~\cite{tzou_2017_schnakenberg}, \cite{schnak2d},
\cite{trinh_2016}, and \cite{chen}). Recently, in \cite{wong} it has
been shown using a hybrid analytical-numerical approach that
peanut-splitting instabilities are subcritical for the 2-D
Schnakenberg, Gray-Scott, and Brusselator models, although the
corresponding issue in a 3-D setting is still an open problem. Our
numerical computations below suggest that peanut-splitting
instabilities for the 3-D GM model in the $D=\varepsilon^2 D_0$ regime
are also subcritical. Moreover, due to the exponentially small
interaction between spots, we also hypothesize that a peanut-splitting
instability triggers a cascade of spot self-replication events that
will eventually pack the domain with identical spots. To explore this
proposed behaviour we use FlexPDE6 \cite{flexpde} to numerically solve
\eqref{eq:pde_gm_3d} in the unit ball with parameters $\tau=1$,
$\varepsilon=0.05$ and $D_0=16\varepsilon^2$, where the initial condition
is a single spot pattern given asymptotically by the solution to
\eqref{eq:core_eps2} with $V(0)=5$. From the bifurcation and stability
plots of Figure \ref{fig:eps2} our parameter values and initial
conditions are in the range where a peanut-splitting instability
occurs. In Figure \ref{fig:splitting} we plot contours of the solution
$v(x,t)$ at various times. We observe that the peanut-splitting
instability triggered between $t=20$ and $t=60$ leads to a
self-replication process resulting in two identical spots at $t=110$. The
peanut-splitting instability is triggered for each of these two spots
and this process repeats, leading to a packing of the
domain with $N=8$ identical spots.

\section{General Gierer-Meinhardt Exponents}\label{sec:gen_gm}

Next, we briefly consider the generalized GM model
\begin{equation}\label{eq:pde_gm_3d_general}
   v_t = \varepsilon^2 \Delta v - v + u^{-q} v^p\,,\quad
   \tau u_t = D \Delta u - u + \varepsilon^{-2} u^{-s} v^m\,,
   \quad x\in\Omega\,; \quad \partial_n v = \partial_n u = 0\,, \quad
   x\in\partial\Omega\,,
\end{equation}
where the GM exponents $(p,q,m,s)$ satisfy the usual conditions $p>1$,
$q>0$, $m>1$, $s\geq 0$, and $\zeta \equiv {mq/(p-1)} - (s+1) > 0$
(cf.~\cite{ward_2003_hopf}). Although this general exponent set leads
to some quantitative differences as compared to the prototypical set
$(p,q,m,s)=(2,1,2,0)$ considered in this paper, many of the qualitative
properties resulting from the properties of $\mu(S)$ in Conjecture
\ref{conjecture}, such as the existence of symmetric quasi-equilibrium spot
patterns in the $D={\mathcal O}(1)$ regime, remain unchanged.

Suppose that \eqref{eq:pde_gm_3d_general} has an $N$-spot
quasi-equilibrium solution with well-separated spots.  Near the
$i^\text{th}$ spot we introduce the inner expansion
$v \sim D^\alpha V_i(y)$, $u \sim D^\beta U_i(y)$, and
$y = \varepsilon^{-1}(x-x_i)$, where
\begin{equation*}
  \Delta V_i - V_i + D^{(p-1)\alpha-q\beta} U_i^{-q} V_i^{p} = 0\,,\qquad
  \Delta U_i - \varepsilon^2 D^{-1}U_i  = -D^{m\alpha - (s+1)\beta - 1} U_i^{-s}V_i^m\,,
  \qquad y\in \R^3 \,.
\end{equation*}
Choosing $\alpha$ and $\beta$ such that $(p-1)\alpha - q\beta = 0$ and
$m\alpha - (s+1)\beta = 1$ we obtain
\begin{equation*}
  \alpha = {\nu/\zeta}\,,\qquad \beta = {1/\zeta} \,,\qquad
  \nu = {q/(p-1)}\,,
\end{equation*}
with which the inner expansion takes the form
$v \sim D^{\nu/\zeta} V(\rho;S_{i\varepsilon})$ and
$u\sim D^{1/\zeta}U(\rho;S_{i\varepsilon})$, where $V(\rho;S)$ and
$U(\rho;S)$ are radially-symmetric solutions to the
$D$\textit{-independent} core problem
\begin{subequations}\label{eq:core_general}
\begin{align}
  & \Delta_\rho V - V + U^{-q} V^p = 0\,,\qquad \Delta_\rho U = -U^{-s} V^m\,,
    \qquad \rho>0\,, \label{eq:core_pde_general}\\
  \partial_\rho V(0) =  \partial_\rho & U(0)  = 0\,,\qquad
  V\longrightarrow 0\quad\text{and} \quad U \sim \mu(S) + {S/\rho}\,,
    \qquad \rho\rightarrow\infty\,. \label{eq:core_bc_general}
\end{align}
\end{subequations}
By using the divergence theorem, we obtain the identity
$S = \int_0^\infty U^{-s} V^m \rho^2 \, d\rho > 0$.

\begin{figure}[t!]
  \begin{center}
	\begin{subfigure}[b]{0.45\textwidth}
          \centering
   		\includegraphics[width=\textwidth,height=4.2cm]{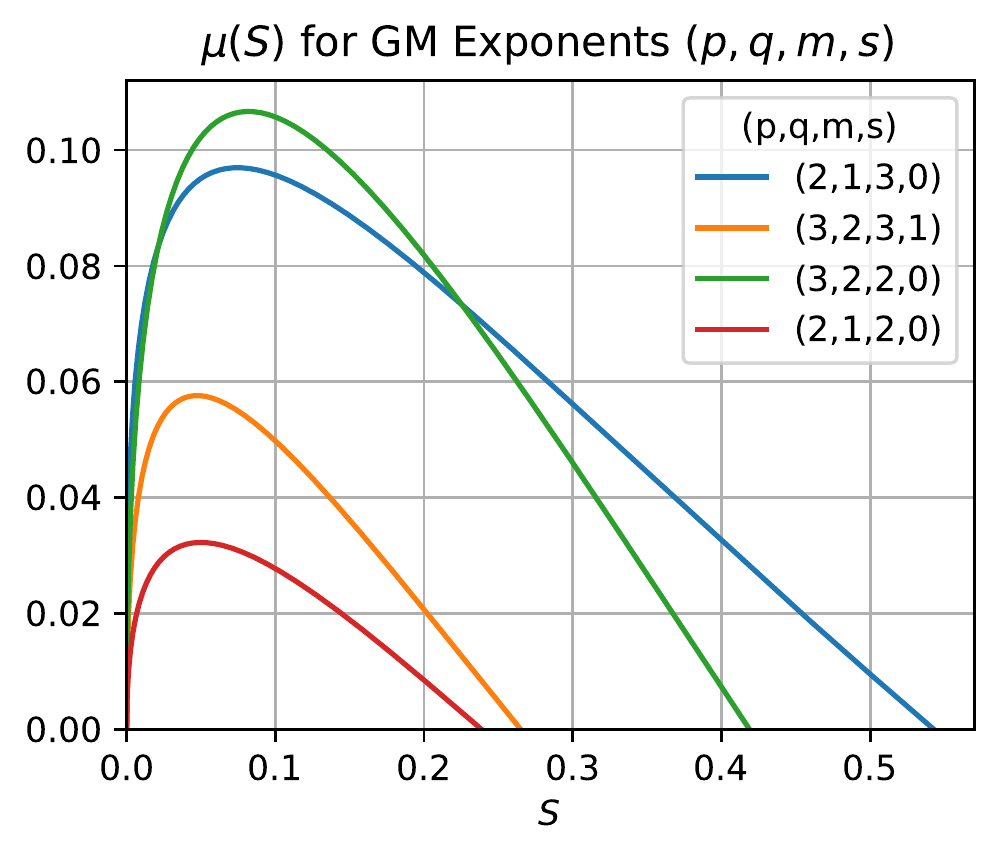}
                \caption{}
                \label{fig:gmgen_pqms}
              \end{subfigure}%
              	\begin{subfigure}[b]{0.45\textwidth}
		\centering
		\includegraphics[width=\textwidth,height=4.2cm]{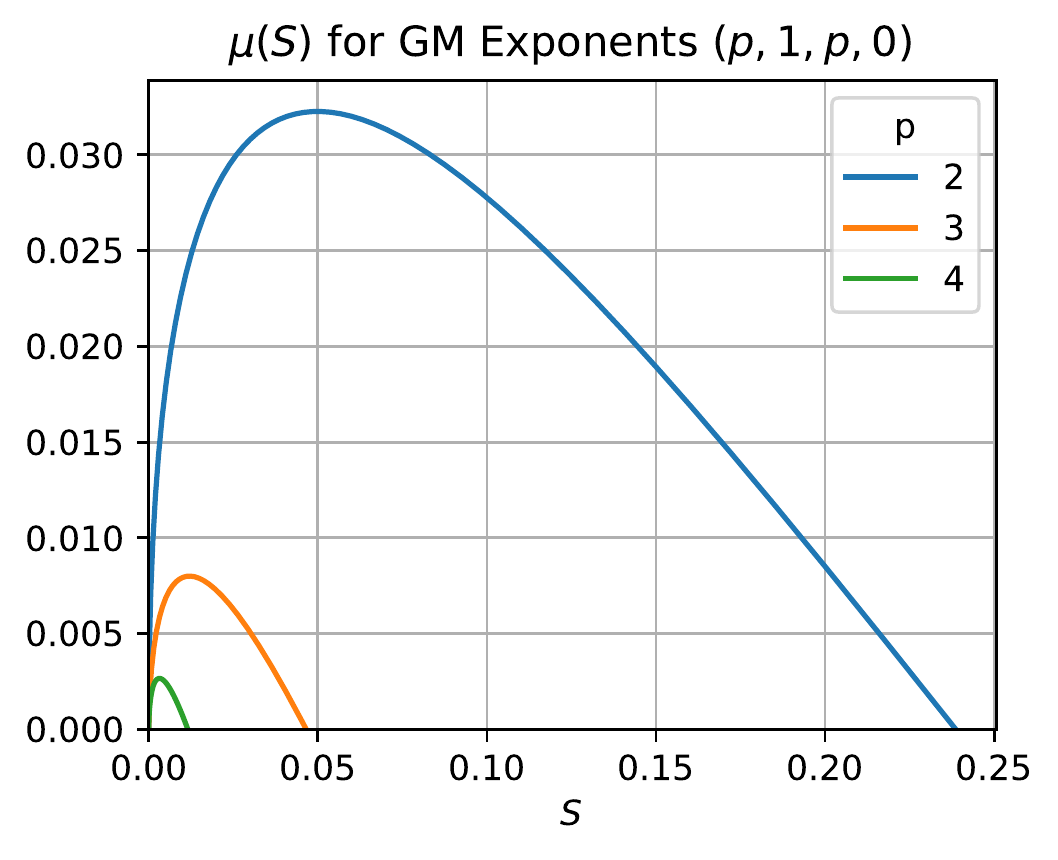}
		 \caption{}
                \label{fig:gmgen_p}
	\end{subfigure}%
	\caption{Left panel: Plot of $\mu(S)$, computed from the
          generalized GM core problem \eqref{eq:core_general}, for the
          indicated exponent sets $(p,q,m,s)$. Right panel: $\mu(S)$
          for exponent sets $(p,1,p,0)$ with $p=2,3,4$. For each set,
          there is a unique $S=S_\star$ for which $\mu(S_\star)=0$. The
          properties of $\mu(S)$ in Conjecture \ref{conjecture}
          for the protypical set $(2,1,2,0)$ still
          hold.}\label{fig:gmgen}
     \end{center}
\end{figure}

By solving the core problem \eqref{eq:core_general} numerically, we
now illustrate that the function $\mu(S)$ retains several of the key
qualitative properties of the exponent set $(p,q,m,s)=(2,1,2,0)$
observed in \S \ref{subsec:core_problem}, which were central to the
analysis in \S \ref{sec:quasi} and \S \ref{sec:stability}. To
path-follow solutions, we proceed as in \S \ref{subsec:core_problem}
by first approximating solutions to \eqref{eq:core_general} for
$S\ll 1$. For $S\ll 1$, we use the identity
$S = \int_0^\infty U^{-s} V^m \rho^2 \, d\rho > 0$ to motivate a small
$S$ scaling law, and from this we readily calculate that
\begin{equation}
  V(\rho;S) \sim \biggl(\frac{S}{b}\biggr)^{\tfrac{\nu}{\zeta+1}} w_c(\rho)\,,
  \quad U(\rho;S) \sim \biggl(\frac{S}{b}\biggr)^{\tfrac{1}{\zeta+1}}\,,\quad
  \mu(S) \sim \biggl(\frac{S}{b}\biggr)^{\tfrac{1}{\zeta+1}}\,, \qquad
   b \equiv \int_0^\infty w_c^m \rho^2 d\rho\,,
\end{equation}
where $w_c>0$ is the radially-symmetric solution of
\begin{equation}\label{eq:homoclinic_equation_general}
  \Delta_\rho w_c - w_c + w_c^p = 0\,,\quad \rho>0\,; \qquad
  \partial_\rho w_c(0) = 0\, ,\qquad w_c\rightarrow 0\quad\text{as}\,\,
  \rho\rightarrow\infty\,.
\end{equation}
With this approximate solution for $S\ll 1$, we proceed as in
\S\ref{subsec:core_problem} to calculate $\mu(S)$ in
\eqref{eq:core_general} for different GM exponent sets by
path-following in $S$. In Figure \ref{fig:gmgen_p} we plot $\mu(S)$
when $(p,q,m,s) = (p,1,p,0)$ with $p=2,3,4$, while a similar plot is
shown in Figure \ref{fig:gmgen_pqms} for other typical exponent sets
in \cite{ward_2003_hopf}. For each set considered, we find that
$\mu(S)$ satisfies the properties in Conjecture
\ref{conjecture}. Finally, to obtain the NAS for the source strengths
we proceed as in \S \ref{quasi:match} to obtain that the outer
solution for the inhibitor field is given by simply replacing $D$ with
$D^{1/\zeta}$ in \eqref{eq:u_outer}. Then, by using the matching
condition
$u\sim D^{1/\zeta}\left(\mu(S_{j\varepsilon}) +
  S_{j\varepsilon}{\varepsilon/|x-x_j|}\right)$ as $x\to x_j$, for
each $j=1,\ldots,N$, we conclude that the NAS \eqref{eq:NAS} still
holds for a general GM exponent set provided that $\mu(S)$ is now
defined by the generalized core problem \eqref{eq:core_general}.

\section{Discussion}\label{sec:discussion}

We have used the method of matched asymptotic expansions to construct
and study the linear stability of $N$-spot quasi-equilibrium solutions
to the 3-D GM model \eqref{eq:pde_gm_3d} in the limit of an
asymptotically small activator diffusivity $\varepsilon\ll 1$. Our key
contribution has been the identification of two distinguished regimes
for the inhibitor diffusivity, the $D={\mathcal O}(1)$ and
$D={\mathcal O}(\varepsilon^{-1})$ regimes, for which we constructed
$N$-spot quasi-equilibrium patterns, analyzed their linear stability,
and derived an ODE system governing their slow spot dynamics. We
determined that in the $D={\mathcal O}(1)$ regime all $N$-spot
patterns are, to leading order in $\varepsilon$, symmetric and
linearly stable on an ${\mathcal O}(1)$ time scale. On the other hand,
in the $D={\mathcal O}(\varepsilon^{-1})$ regime we found the
existence of both symmetric and asymmetric $N$-spot patterns. However,
we demonstrated that all asymmetric patterns are unstable on an
${\mathcal O}(1)$ time scale, while for the symmetric patterns we
calculated Hopf and competition instability thresholds. These GM
results are related to those in \cite{tzou_2017_schnakenberg} for the
3-D singularly perturbed Schnakenberg model, with one of the key new
features being the emergence of \textit{two} distinguished limits, and
in particular the existence of localized solutions in the
$D={\mathcal O}(1)$ regime for the GM model. For $D={\mathcal O}(1)$,
concentration behavior for the Schnakenberg model as
$\varepsilon\to 0$ is no longer at discrete points typical of spot
patterns, but instead appears to occur on higher co-dimension
structures such as thin sheets and tubes in 3-D (cf.~\cite{tzou}). For
the GM model, we illustrated the onset of both Hopf and competition
instabilities by numerically solving the full GM PDE system using the
finite element software FlexPDE6 \cite{flexpde}. We have also
considered the weak-interaction regime
$D={\mathcal O}(\varepsilon^2)$, where we used a hybrid
analytical-numerical approach to calculate steady-state solutions and
determine their linear stability properties. In this small $D$ regime
we found that spot patterns are susceptible to peanut-splitting
instabilities. Finally, using FlexPDE6 we illustrated how the
weak-interaction between spots together with the peanut-splitting
instability leads to a cascade of spot self-replication events.

\begin{figure}
	\centering
	\begin{subfigure}[b]{0.333\textwidth}
		\centering
		\includegraphics[width=\textwidth,height=4.2cm]{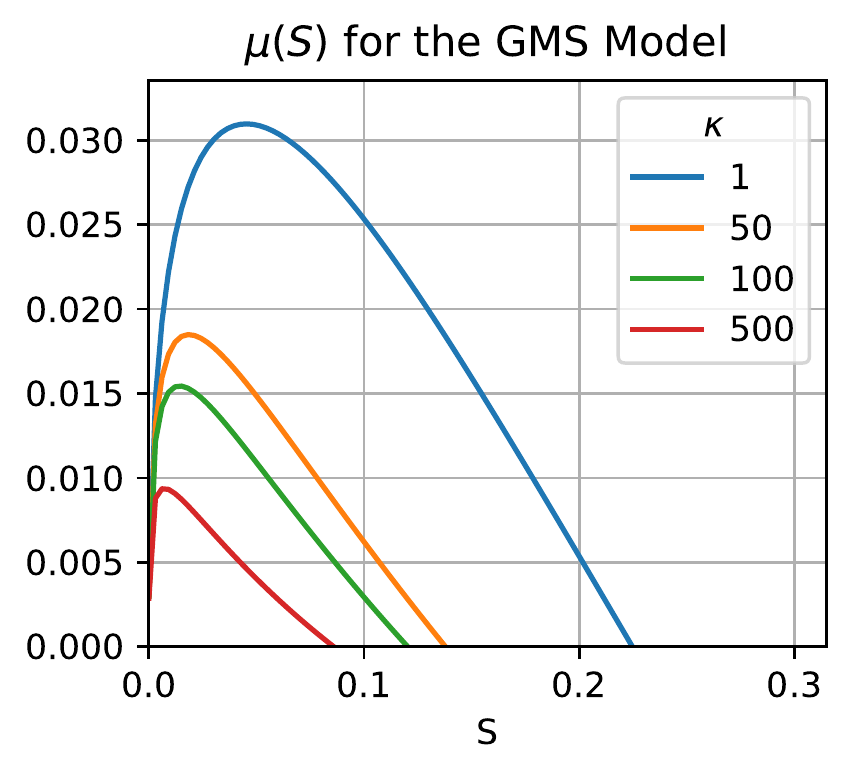}
		 \caption{}
                \label{fig:gms_model}
	\end{subfigure}%
	\begin{subfigure}[b]{0.333\textwidth}
		\centering
		\includegraphics[width=\textwidth,height=4.2cm]{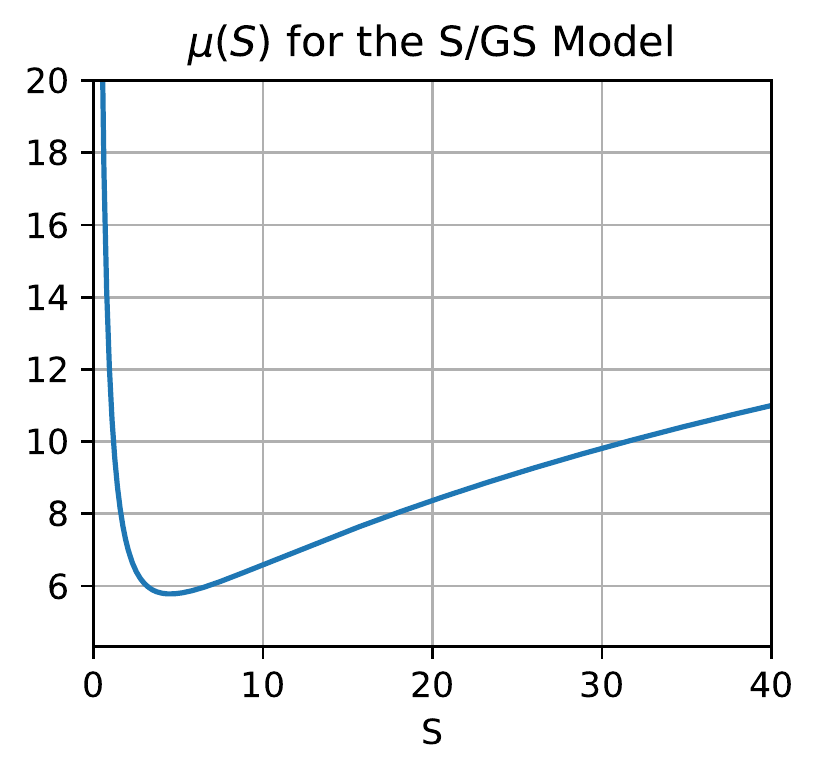}
		 \caption{}
                \label{fig:sgs_model}
	\end{subfigure}%
	\begin{subfigure}[b]{0.333\textwidth}
		\centering
		\includegraphics[width=\textwidth,height=4.2cm]{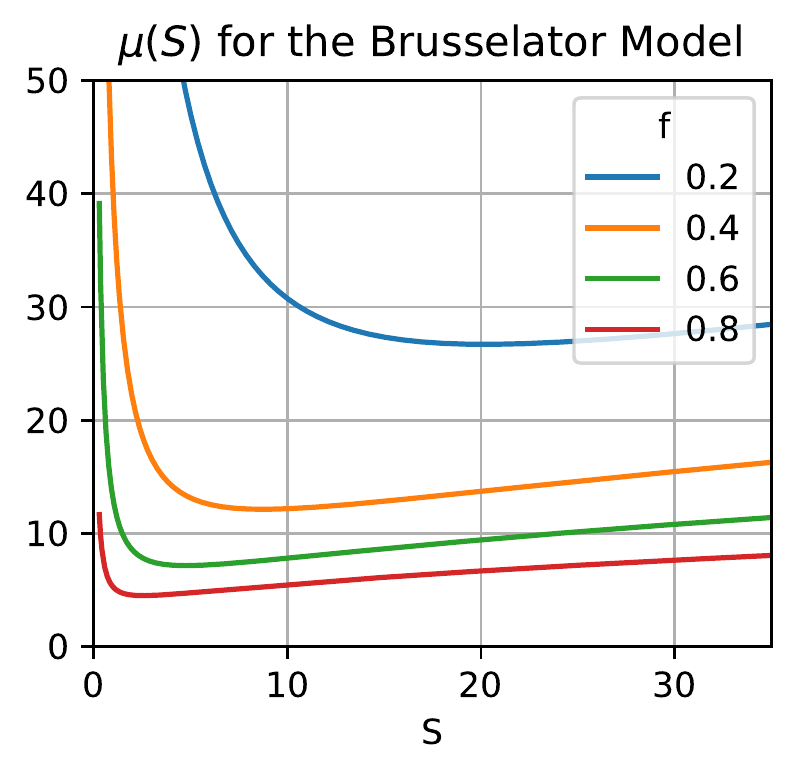}
		 \caption{}
                \label{fig:b_model}
	\end{subfigure}%
	\caption{Plots of the far-field constant behaviour for the (a)
          Gierer-Meinhardt with saturation, (b) Schnakenberg or
          Gray-Scott, and (c) Brusselator models. See Table
          \ref{tbl:core_problems} for the explicit form of the
          kinetics $F(v,u)$ and $G(v,u)$ for each
          model. A zero-crossing of $\mu(S)$ at some $S>0$ occurs
          only for the GMS model.}\label{fig:mu_model_plots}
\end{figure}

We conclude by highlighting directions for future work and open
problems.  First, although we have provided numerical evidence for the
properties of $\mu(S)$ highlighted in Conjecture \ref{conjecture}, a
rigorous proof remains to be found. In particular, we believe that it
would be significant contribution to rigorously prove the existence
and uniqueness of the \textit{ground state} solution to the core
problem \eqref{eq:core}, which we numerically calculated when
$S=S_\star$. A broader and more ambitious future direction is to
characterize the reaction kinetics $F(V,U)$ and $G(V,U)$ for which the
core problem
\begin{equation}\label{new_eq:core_general}
  \Delta_\rho V + F(V,U) = 0,\qquad \Delta_\rho U + G(V,U) = 0,\qquad\text{in}
  \,\,\,   \rho>0\,,
\end{equation}
admits a radially-symmetric ground state solution for which
$V\rightarrow 0$ exponentially and $U = {\mathcal O}(\rho^{-1})$ as
$\rho\rightarrow\infty$. The existence of such a ground state plays a
key role in determining the regimes of $D$ for which localized
solutions can be constructed. For example, in the study of the 3-D
singularly perturbed Schnakenberg model it was found that the core
problem does not admit such a solution and as a result localized spot
solutions could not be constructed in the $D={\mathcal O}(1)$ regime
(cf.~\cite{tzou_2017_schnakenberg}). To further motivate such an
investigation of \eqref{new_eq:core_general} we extend our numerical
method from \S\ref{subsec:core_problem} to calculate and plot in
Figure \ref{fig:mu_model_plots} the far-field constant $\mu(S)$ for
the core problems associated with the GM model with saturation (GMS),
the Schnakenberg/Gray-Scott (S/GS) model, and the Brusselator (B)
model (see Table \ref{tbl:core_problems} for more details). Note that
for the GMS model we can find values of $S_\star$ such that
$\mu(S_\star)=0$, but such a zero-crossing does not appear to occur
for the (S/GS) and (B) models. As a consequence, for these three
specific RD systems, localized spot patterns in the
$D={\mathcal O}(1)$ regime should only occur for the GMS model. Additionally,
understanding how properties of $\mu(S)$, such as convexity and
positiveness, are inherited from the reaction kinetics would be a
significant contribution.  In this direction, it would be interesting
to try extend the rigorous numerics methodology of \cite{lessard} to
try to establish Conjecture \ref{conjecture}.

\begin{table}[b!]
  \begin{center}
	\begin{tabular}{|c|c|c|c|}
		\hline
		RD Model & $F(V,U)$ & $G(V,U)$ & Decay behavior \\ \hline \\[-2ex]
		Gierer-Meinhardt with Saturation (GMS) & $-V + \frac{V^2}{U(1+\kappa U^2)}$ & $V^2$ & $U\sim \mu(S) + S/\rho$ \\[1.5ex] \hline \\[-2ex]
		Schnakenberg or Gray-Scott (S/GS) & $-V+V^2U$ & $-V^2U$ &  $U\sim\mu(S)-S/\rho$ \\[1.5ex] \hline \\[-2ex]
		Brusselator (B) & $-V + fV^2 U$ & $V - V^2 U$ & $U\sim\mu(S)-S/\rho$ \\[1.5ex] \hline
	\end{tabular}
	\caption{Core problems and far-field inhibitor
          behaviour for some common RD systems.}\label{tbl:core_problems}
        \end{center}
\end{table}

\section*{Acknowledgments}
Daniel Gomez was supported by an NSERC Doctoral Fellowship. Michael Ward
and Juncheng Wei gratefully acknowledge the support of the NSERC Discovery
Grant Program.

\addcontentsline{toc}{section}{References}
\bibliographystyle{abbrv}
\bibliography{bibliography}

\end{document}